\documentclass[reprint,amsmath,amssymb,mathtools,aps,pra]{revtex4-2}

\usepackage{graphicx}
\usepackage{subcaption}
\usepackage{dcolumn}
\usepackage{bm}
\usepackage{lipsum}
\usepackage{xcolor}
\newcommand{\ii}{\text{i}}

\begin{document}


\title{The dynamics of a single trapped ion in a high density
media: a stochastic approach}

\author{Mateo Londoño}
 \email{mateo.londono@correounivalle.edu.co}
\author{Javier Madroñero}%
\affiliation{%
Departamento de Física, Universidad del Valle, A.A. 25360, Cali, Colombia 
}%
\author{Jesús Pérez-Ríos}%
 
\affiliation{Department of Physics and Astronomy, Stony Brook University, Stony Brook, NY 11794, USA 
}%

\date{\today}

\begin{abstract}
Based on the Langevin equation, a stochastic formulation is implemented to describe the dynamics of a trapped ion in a bath of ultracold atoms, including an excess of micromotion. The ion dynamics is described following a hybrid analytical-numerical approach in which the ion is treated as a classical impurity in a thermal bath. As a result, the ion energy's time evolution and distribution are derived from studying the sympathetic cooling process. Furthermore, the ion dynamics under different stochastic noise terms is also considered to gain information on the bath properties' role in the system's energy transfer processes. Finally, the results obtained from this formulation are contrasted with those obtained with a more traditional Monte Carlo approach.  

\end{abstract}

--\maketitle


\section{\label{sec:introcduction} Introduction}

In recent years it has been possible to combine optical traps and Paul traps in a single experimental setup to obtain cold atom-ion hybrid systems~\cite{ion-atom1}, combining two emerging research fields: ultracold atoms and cold ions~\cite{ion-atom2}. These hybrid systems are expected to incorporate the essential advantages of the two components representing a new promising scenario to study fundamental aspects of matter and light-matter interaction and to develop quantum technologies \cite{reviewhybrid}. Hybrid ion-atom systems allow us to study charged-neutral interactions in a controllable way, thus, revolutionizing physical chemistry and paving the way to a new field of research: cold chemistry~\cite{J_perez_book}. Additionally, atom-ion hybrid systems offer a platform for studying the impurity physics of a charged particle in a neutral sea. This problem is usually approached by a many-body~\cite{MBImpur1, MBImpur2} or a few-body perspective~\cite{FBImpur}. 

 Most of the prospects and applications of hybrid atom-ion systems rely on reaching the quantum $s$-wave regime \cite{reviewhybrid}. Ultracold atomic gases can be readily prepared in the $100\,\text{nK}$ regime \cite{BEC1, Fermi1}. Then, it is necessary to use sympathetic cooling to cold down ions by bringing in contact ions with an ultracold gas (buffer gas). However, the time dependence of the electric field in Paul traps limits the temperature of the trapped ions in a buffer gas \cite{F.Major_collision}. Although, using atom-ion combinations with a large mass ratio \cite{Cetina} such as $^{6}$Li-Yb$^{+}$  can alleviate this effect. In general, atom-ion dynamics is studied from three different approaches: a Monte Carlo \cite{Zipkes}, molecular dynamics \cite{H.Furst} and few-body {\it ab initio} \cite{JogerLi-Yb,BellaouiniSr-Na, SaitoLi-Ca}. However, there is an alternative perspective to atom-ion systems in which most of the degrees of freedom of the bath are integrated out, yielding an effective ion-bath interaction.

 This work studies the dynamics of a single ion in a buffer gas from a stochastic approach. In particular, instead of considering all degrees of freedom of the bath, these are substituted by an effective stochastic force. Such a force can be modeled with different types of noise terms, which in principle allows exploring the role of the bath properties in the ion's dynamic and cooling processes via solving the Langevin equation —giving the possibility of gaining new insights into the mechanisms of energy transfer between the ion the bath. The paper is structured as follows. The Sec. \ref{langevin_model} is devoted to explaining the model based on the Langevin equation (LE) and the generalized Langevin equation (GLE) for the ion in the atomic bath, and Sec. \ref{white_noise_sec} and \ref{colored_noise_sec} present the main results of this formulation. In Sec. \ref{molecular_dynamics_sec}, the more traditional hard spheres molecular dynamics methodology is introduced, and the results are contrasted with those of the stochastic formulation. Finally, Sec. \ref{conclusions} summarizes the main results of the work and presents some perspectives.

\section{\label{langevin_model} Langevin equation model}

It is well-known that the motion of a single ion in a Paul trap is described by the Mathieu equation \cite{J_perez_book}

\begin{equation}\label{Mathieu_equation}
\frac{\text{d}^{2}r_{j}}{\text{d}t^{2}} + \frac{\Omega^2}{4}(a_{j} + 2q_{j}\cos(\Omega t))r_{j} = \frac{F_{\rm{mm},j}}{m},
\end{equation}
where $j=(x,y,z)$, $\Omega$ is the trap frequency, and the coefficients $a_{j}, q_{j}$ depends on the trap parameters and the ion mass, $m$. $F_{\rm{mm},j}$ is the force due to possible external fields or \textit{excess micromotion} sources, such as stray electric fields or phase difference between the AC potentials applied to the electrodes~\cite{Berkeland}. The homogeneous form of Eq. (\ref{Mathieu_equation}) admits solutions via Floquet  theorem \cite{H.Dehmelt68}. In particular, in the limit of $a_{j} \ll q_{j}^{2}/2 \ll 1$ and making use of the adiabatic approximation, the approximate ion's position is given by

\begin{equation*}
\begin{split}
r_{j}(t) &\approx A_{j}\cos(\omega_{j} t + \phi)\Big( 1+\frac{q_{j}}{2}\cos(\Omega t)\Big)\\
&= A_{j}\cos(\omega_{j} t + \phi) + A_{j}\frac{q_{j}}{2}\cos(\omega_{j} t + \phi)\cos(\Omega t),   
\end{split}
\end{equation*}
where constants A$_{j}$ and $\phi_{j}$ depend on the initial conditions. The adiabatic or pseudopotential approximation consists of separating the slow secular motion, driven by the secular frequency $\omega_{j} = \frac{\Omega}{2}\sqrt{a_{j} + \frac{q_{j}^{2}}{2}}$, from the fast superimposed micromotion driven by frequencies $\Omega\pm\omega_{j}$.

The dynamics of a trapped ion in contact with a buffer gas is more involved due to its collisions with the gas particles. In this scenario, one possible way to treat such dynamics is assuming that the ion behaves as a trapped Brownian particle in a given bath. As a result, it is possible to simulate the ion's dynamics as a Langevin stochastic process. In particular, in the case of an ion in a linear Paul trap, the degrees of freedom are decoupled, leading to the following stochastic differential equation for the motion for the  $j$-th component

\begin{equation}\label{GLE}
\begin{split}
\frac{\text{d}^{2}r_{j}}{\text{d}t^{2}} + \frac{1}{m}\int_{0}^{t}\Gamma(t-t^{\prime})v_{j}(t^{\prime})dt^{\prime}+&\frac{\Omega^{2}}{4}[a_{j} + 2q_{j}\cos(\Omega t)]r_{j}\\
&= \frac{F_{\rm{mm},j}}{m}  + \frac{\zeta(t)}{m},
\end{split}
\end{equation}

\noindent
where the new terms in comparison to Eq. (\ref{Mathieu_equation}), $ \int_{0}^{t}\Gamma(t-t^{\prime})v_{j}(t^{\prime})dt^{\prime}$ and $\zeta(t)$, represent forces arising from the interaction between the ion and the bath. The former represents a friction force coming from the few-body physics of atom-ion interactions, whereas the latter represents a stochastic force. 

$\zeta(t)$ is a random force resulting from the thermal fluctuations associated with the bath after integrating out most of their degrees of freedom. The random force is fully determined by its statistical properties as

\begin{equation}\label{CN}
\langle \zeta_{c}(t) \rangle = 0 \,\,\,\,\,\, \text{and} \,\,\,\,\,\, \langle \zeta_{c}(t)\zeta_{c}(s) \rangle = C(t-s),
\end{equation}
where $C(t-s)$ is the force correlation function at two different times, $t$ and $s$. The friction and the stochastic forces are intimately related via the fluctuation-dissipation theorem (FDT) as \footnote{It is worth noticing that even in the presence of an external field, the fluctuation-dissipation theorem holds since the external field only affects the Brownian particle.}

\begin{equation}\label{FDT}
    \Gamma(t-t^{\prime}) = \frac{1}{2k_{\text{B}}T}C(t-t^{\prime}),
\end{equation}
where $k_\text{B}$ is the Boltzmann constant, and $T$ is the temperature of the bath. 

In the following, we use two distinct models for the stochastic force, namely a white and colored noise model, to study the behavior of a trapped ion interacting with an atomic bath. In addition, the results are contrasted with results from Monte Carlo simulations. 

\section{White noise bath}
\label{white_noise_sec}

Assuming that the stochastic force represents a white noise, we find 

\begin{equation}\label{WN}
\langle \zeta(t) \rangle = 0 \,\,\,\,\,\, \text{and} \,\,\,\,\,\, \langle \zeta(t)\zeta(s) \rangle = D\delta(t-s),
\end{equation}
where $\delta(x)$ stands for the Dirac delta function of argument $x$, and $D$ is the strength of the noise. Eq. (\ref{WN}) establishes that the force at two different times is uncorrelated and hence, the force acting on the ion only depends on the actual state of the bath. That is, a white noise leads to Markovian dynamics of the ion in a bath. 

In this scenario, the FDT yields

\begin{equation}
    \Gamma(t-t^{\prime}) = \frac{D}{2k_{\text{B}}T}\delta(t-t^{\prime}),
\end{equation}
and the ion's dynamics is given by the following Langevin equation

\begin{equation}\label{LE}
\begin{split}
\frac{\text{d}^{2}r_{j}}{\text{d}t^{2}} + \frac{\gamma}{m}\frac{\text{d}r_{j}}{\text{d}t}+ &\frac{\Omega^{2}_{\rm{RF}}}{4}[a_{j} + 2q_{j}\cos(\Omega_{\rm{RF}}t)]r_{j}
= \frac{F_{\rm{mm},j}}{m} + \frac{\zeta(t)}{m},
\end{split}
\end{equation}
where we have introduced the friction coefficient defined as $\gamma = \frac{D}{2k_{\text{B}}T}$.

\subsection{Dynamics of the ion}

\begin{figure}
\begin{subfigure}[h]{0.98\linewidth}
\includegraphics[width=0.9\linewidth]{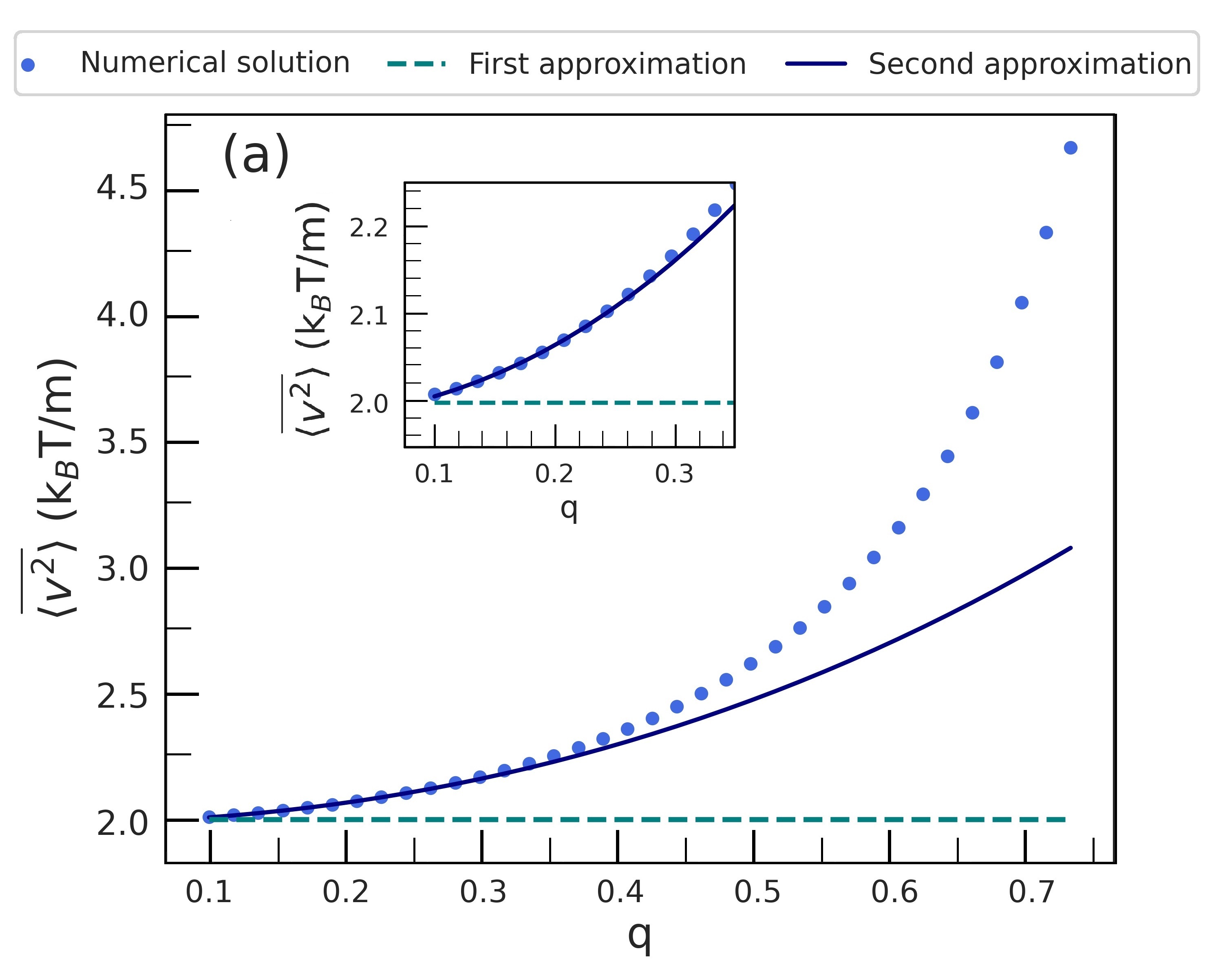}
\label{sqv_vs_q_no_field}
\end{subfigure}
\begin{subfigure}[h]{0.98\linewidth}
\includegraphics[width=\linewidth]{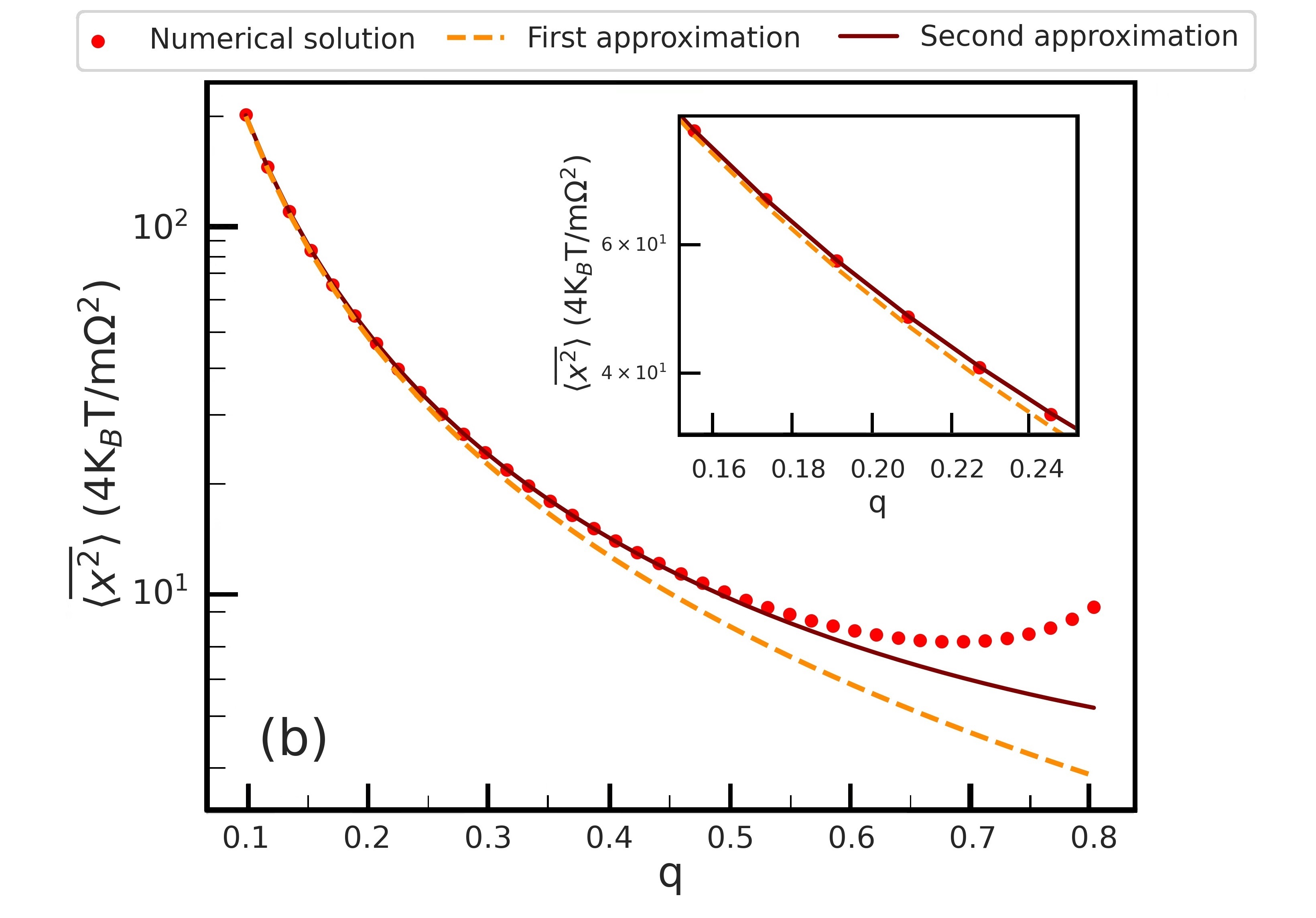}
\end{subfigure}%
\caption{Variation of the mean square values with respect to the $q$ parameter. The trap parameter was kept as $a=$-8$\times10^{-6}$ and $\Omega = 2\pi\times10^{6}$ Hz. The points are the numerical calculation, the solid line is the first approximation described by Eq. (\ref{time_avg_msr_1}) and the second  approximation is given by Eq. (\ref{time_avg_msr_2}). Panels (a) and (b) show $\overline{{\langle v^{2} \rangle}}$ vs $q$ and $\overline{{\langle x^{2} \rangle}}$ vs $q$ respectively. Aditionally, the calculations were performed considering a  $^{171}$Yb$^{+}$ ion in an ultracold cloud of $^{6}$Li  }
\label{fig1}
\end{figure}
Here, we assume a linear Paul trap configuration such that $q_{x} = -q_{y}$, $q_{z}=0$, $a_{x}=a_{y}= \frac {a_{z}}{2}$, satisfying $a_{j} \ll q_{j}^{2}/2 \ll 1$. In addition, we  explicitly include the additional term  $\frac{F_{\text{mm},j}}{m}= \frac{e E_{j}}{m} = \tilde{E}_{j} $ ($j = {x,y}$) resulting from a radial stray  electric field. The few-body physics regarding atom-ion interaction is encapsulated in the diffusion coefficient, which is obtained through the Chapman-Enskog approximation~\cite{kinetic_theory_book} fed with the thermally averaged diffusion cross section, as it is described on the appendix~\ref{appendixA}. Next, in virtue of the FDT, it is possible to calculate the friction coefficient from the diffusion coefficient, and solve the Langevin equation leading to the following equations for the mean position and velocity of the ion

\begin{equation}\label{expected_white}
\begin{split}
\frac{\text{d}\langle v_{j}\rangle}{\text{d}t} + \gamma^{\prime}\langle v_{j}\rangle+ \frac{\Omega^{2}}{4}&[a_{j} + 2q_{j}\cos(\Omega t)]\langle r_{j}\rangle
= \tilde{E}_{j},\\
& \frac{\text{d}\langle r_{j}\rangle}{\text{d}t}= \langle v_{j} \rangle,
\end{split}
\end{equation}
\noindent
where we have made $\frac{\gamma}{m} = \gamma^{\prime}$ to simplify the notation. Hence, the time-evolution of the mean square values of velocity and position, and the cross-correlations read as

\begin{align}\label{correlations_Eq}
\frac{\text{d}}{\text{d}t}
\mathop{\begin{pmatrix}
    \langle r_{j}^{2} \rangle \\
    \langle v_{j}^{2} \rangle \\
    \langle r_{j}v_{i} \rangle
\end{pmatrix}}
=
\textbf{M}
\mathop{\begin{pmatrix}
    \langle r_{j}^{2} \rangle \\
    \langle v_{j}^{2} \rangle \\
    \langle r_{j}v_{i} \rangle
\end{pmatrix}} + \tilde{E}_{j}
\mathop{\begin{pmatrix}
    0 \\
    \langle v_{j} \rangle \\
    \langle r_{j} \rangle
\end{pmatrix}},
\end{align}
with \textbf{M} given by
\begin{align*}
\mathop{\begin{pmatrix}
    0 & 0 & 2 \\
    0 & -2\gamma^{\prime} & -\frac{\Omega^{2}}{2}[a_{j} + 2q_{j}\cos(\Omega t)] \\
    -\frac{\Omega^{2}}{4}[a_{j} + 2q_{j}\cos(\Omega t)] & 1 & -\gamma^{\prime} \\
\end{pmatrix}} 
\end{align*}
\noindent

\begin{figure}[h]
\centering
\includegraphics[width=0.44\textwidth]{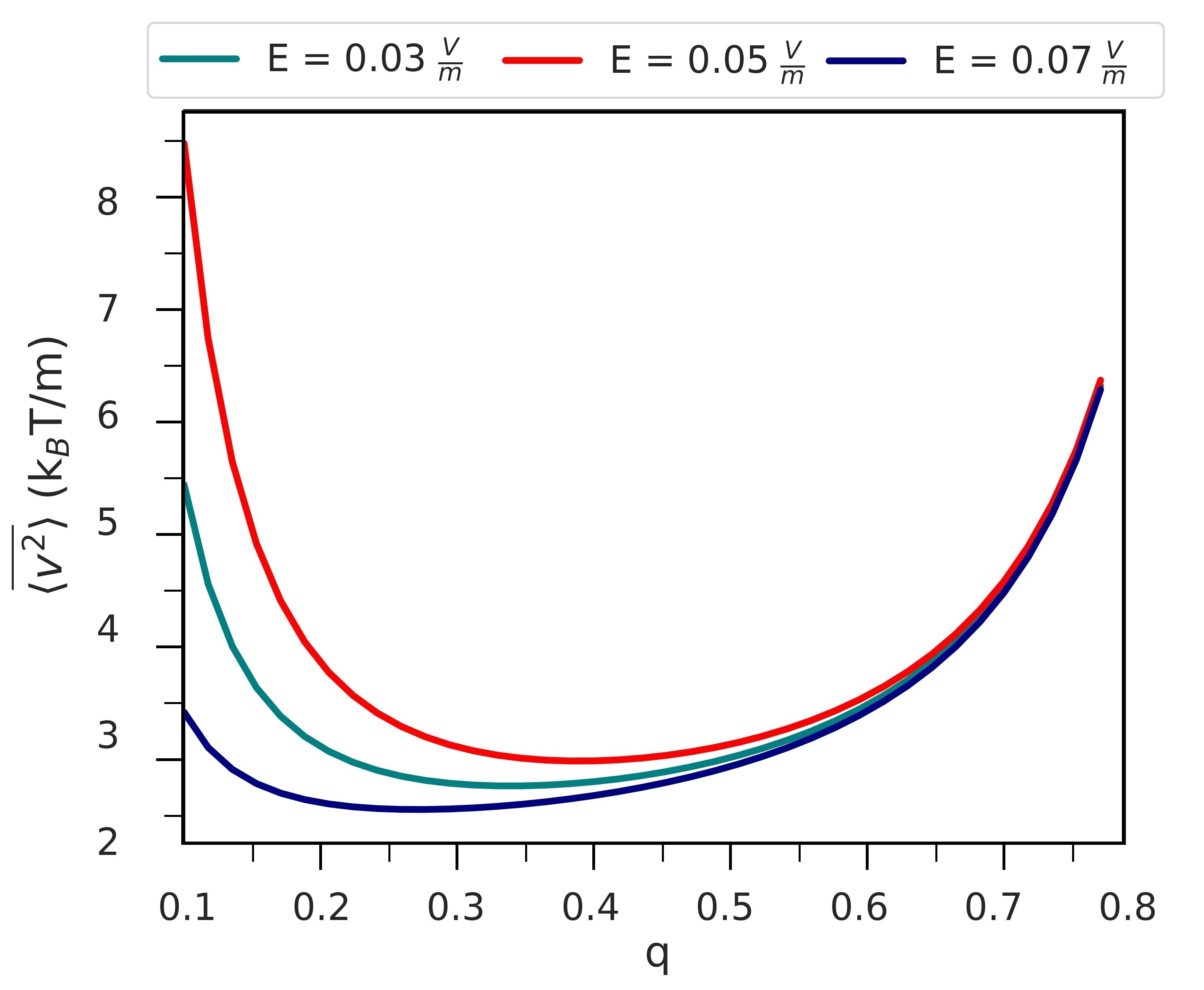}
\caption{Variation of the time average mean square velocity  with respect the $q$ parameter for different stray fields. The trap parameters were fixed to ($a= -8\times10^{-6}$, $\Omega=2\pi \times10^{6}\,$Hz). The curves have been obtained numerically considering a $^{171}$Yb$^{+}$ -$^{6}$Li mixture}  
\label{fig2}
\end{figure}

Eq.~(\ref{correlations_Eq}) is solved numerically using a fourth-order Runge-Kutta method for one of the $j$-components~\footnote{Note that in a Linear Paul trap the dynamics in each coordinate is independent of each other.}, and the results are shown in Figs.~\ref{fig1} and \ref{fig2}. Fig.~\ref{fig1} displays the time-averaged mean square position and velocity, $\overline{\langle x^{2} \rangle} $ and $\overline{\langle {v^{2}} \rangle}$, respectively, as a function of $q_{j}$, keeping $a$ and $\Omega$ fixed in the absence of excess micromotion ($ \tilde{E}_{j}=0$). In this figure, it is noticed that the larger the value of $q$, the smaller the mean square position of the ion is. On the contrary, the average square velocity of the ion increases with $q$. To further understand the observed behavior, we have conducted a theoretical analysis of the average mean square position and velocity of the ion based on a continued-fraction expansion following Ref.~\cite{blatt1986}. In particular, we find that $\langle r_{j}^{2} \rangle$, denoted here as $\sigma_{rr}$ for simplicity, satisfies 

\begin{equation}\label{x_var_equation}
\begin{split}
\dddot{\sigma}_{rr}  = & -3\gamma^{\prime}\ddot{\sigma}_{rr} - \Big(\frac{\Omega^{2}}{2}(a_{j} + 2q_{j}\cos(\Omega)) + 2\gamma^{\prime 2}\Big)\dot{\sigma}_{rr}\\
& -\big( \gamma^{\prime}\Omega^{2}     \big (a_{j} + 2q_{j}\cos(\Omega))) \sigma_{rr} + \frac{2D}{m} \\
& + 4\tilde{E}\gamma^{\prime} \langle r_{j}\rangle + 4\tilde{E}\langle v_{j}\rangle,
\end{split}
\end{equation}
that, in the long time regime, can be solved via the Fourier expansion method using the following ansatz


\begin{equation*}
    \mathop{\begin{pmatrix}
    \langle x^{2} \rangle \\
    \langle v^{2} \rangle \\
    \langle xv \rangle
\end{pmatrix}}
=
\mathop{\begin{pmatrix}
    \sum_{n}x_{n}e^{-in\Omega t} \\
    \sum_{n}v_{n}e^{-in\Omega t} \\
     \sum_{n}c_{n}e^{-in\Omega t} 
\end{pmatrix}}.
\end{equation*}
In the case of $t\gg 1/\gamma^{\prime}$, i.e., a quasi-stationary state is reached and Eq.~(\ref{x_var_equation}), yields  
\begin{equation}
\label{eq11}
\begin{split}
 Q_{n}x_{n} &+ Q_{n}^{-}x_{n-1} +Q_{n}^{+}x_{n+1} =\\ 
& 2\Big(
\frac{D}{m} + \frac{2\tilde{E}_{x}^{2}\gamma^{\prime}}{\omega^{2}} \Big)\delta_{n,0} + \Big(
\frac{\tilde{E}_{x}^{2}q\gamma^{\prime}}{\omega^{2}} \pm \frac{\tilde{E}_{x}^{2}q\Omega}{\omega^{2}}\ii
\Big)\delta_{n,\pm 1},
\end{split}  
\end{equation}
with 
\begin{equation*}
\begin{split}
& Q_{n} = -\ii n^{3}\Omega^{3} - 3\gamma^{\prime} n^{2}\Omega^{2}+ 2(\gamma^{\prime 2} + a\Omega^{2})n\Omega \ii + \gamma^{\prime}\Omega^{2}a\\
&Q_{n}^{\pm} = \frac{q\Omega^{2}}{2}(\Omega \ii(2n\pm 1) + 2\gamma^{\prime}).
\end{split}
\end{equation*}

In the absence of excess micromotion, Eq. (\ref{eq11}) simplifies to $x_{n+1} = S_{n}^{+}x_{n}$ for $n\geq 0$ and $x_{n-1} = S_{n}^{-}x_{n}$ for $n\leq 0$ with 
\begin{equation}\label{S_coefficients}
S_{n}^{\pm} = - \frac{Q_{n\pm1}^{\mp}} {Q_{n\pm1} + Q_{n\pm1}^{\pm}S_{n\pm1}^{\pm} }.
\end{equation}
Next, assuming $\frac{\gamma^{\prime 2}}{\Omega^{2}}\ll1$, the time independent component of the Fourier expansion for the mean squared position is given by
\begin{equation}
\label{time_avg_msr_1}
    \overline{\langle {x^{2}} \rangle} = \frac{2D/m}{Q_{0} + 2\text{Re}(Q_{0}^{+}S_{0}^{+})}\approx \frac{k_{\rm{B}}T}{m\omega^{2}},
\end{equation}
and the velocity and cross-correlation contributions are given by 
\begin{equation*}
\begin{split}
    v_{n} = (\ii n\Omega + \gamma^{\prime})c_{n} + \frac{\Omega^{2}a}{4} x_{n} + & \frac{\Omega^{2}q}{4}(x_{n-1} + x_{n+1})\\
    c_{n} = \frac{\ii n\Omega x_{n}}{2}.
\end{split}
\end{equation*}
In the adiabatic limit $c_{0} = \overline{\langle xv \rangle} = 0$ and hence we find

\begin{equation}
\label{time_avg_msr_1_bis}
\overline{\langle {v^{2}}} \rangle \approx \frac{2k_{\rm{B}}T}{m},
\end{equation}
leading to $\overline{\langle E_{k} \rangle} =\frac{1}{2}m\overline{\langle {v^{2}} \rangle} \approx k_{\text{B}}T$ for the spatial degrees of freedom with an RF field, which is a well-known result. The results from Eqs. (\ref{time_avg_msr_1}) and (\ref{time_avg_msr_1_bis}) are shown as dashed lines in Fig.\ref{fig1}, where it is noticed that they can only properly describe the ion dynamics for $q\lesssim 0.1$.

\begin{table*}[ht]
\caption{Leading contributions of the Fourier expansion of the square position and velocity in the pseudopotential regime.}
\centering
\label{fourier_terms}
\begin{tabular}{p{0.25\linewidth}p{0.25\linewidth}p{0.25\linewidth}}
\hline
 $n$ & x$_{n}$ & v$_{n}$\\
\hline
$0$ & $\frac{k_{\text{B}}T}{m\omega^{2}}$ & $\frac{k_{\text{B}}T}{m}$\\

1 & $\Big( \frac{q}{2} + \frac{q\gamma^{\prime}}{2\Omega}\ii\Big)x_{0}$ & $\frac{-q\gamma^{\prime2}}{4}x_{0}$\\

2 & $\Big( \frac{3}{32}q^{2} + \frac{3}{64}\frac{\gamma^{\prime}}{\Omega }q^{2}\ii \Big)x_{0}$  & $\Big( -\frac{25}{184}q^{2}\Omega^{2} + \frac{113}{368}\gamma^{\prime}\Omega q^{2}\ii\Big)x_{0}$\\

3 &  $\Big( \frac{5}{414}q^{3} - \frac{181}{13248}\frac{\gamma^{\prime}}{\Omega}q^{3}\ii \Big)x_{0}$ & $\Big( -\frac{1}{64}q^{3}\Omega^{2} + \frac{3}{64}\gamma^{\prime}\Omega q^{3}\ii\Big)x_{0}$\\
\hline
\end{tabular}
\end{table*}

Here, we go beyond previous studies by including higher-order contributions up to the third harmonic, i.e., $n=3$, and the results are shown in Table \ref{fourier_terms}. We find the following time-averaged expression for the mean squared position and velocity are 

\begin{equation}
\label{time_avg_msr_2}
    \begin{split}
         &\overline{\langle x^{2} \rangle} \approx  \frac{k_{\text{B}}T}{m\omega^{2}}\bigg[ \frac{1+\frac{27}{50}q^{2}}{1-\frac{1}{5
         }q^{2}}\bigg]\\
         &\overline{\langle {v^{2}} \rangle} \approx \frac{k_{\text{B}}T}{m\omega^{2}}\bigg[ \frac{a\Omega^{2}}{4}\bigg( \frac{1+\frac{14}{25}q^{2}}{1-\frac{24}{31
         }q^{2}}\bigg) + \frac{q^{2}\Omega^{2}}{4}\bigg( \frac{1+\frac{20}{25}q^{2}}{1-\frac{7}{50
         }q^{2}}\bigg) \bigg],
    \end{split}
\end{equation}

\noindent
which are displayed as the solid line in Fig.~\ref{fig1}. As a result,  the average position of the ion depends quadratically on $q$, whereas its kinetic energy shows a more involved behavior. The approximation given by Eq.~(\ref{time_avg_msr_2}) adequately describes the ion dynamics for $q\lesssim 0.3$.

\begin{figure}
\begin{subfigure}[h]{0.93\linewidth}
\includegraphics[width=\linewidth]{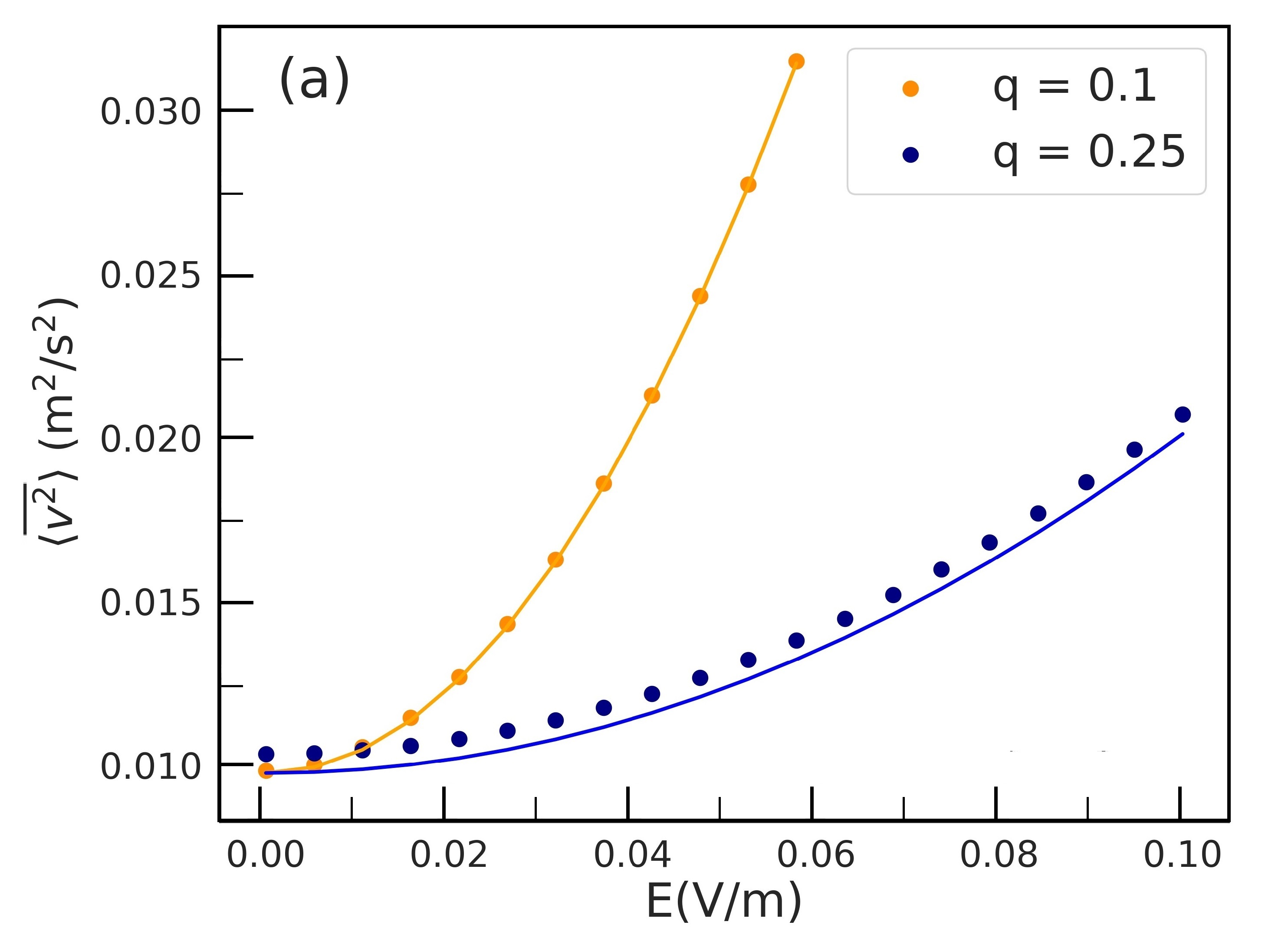}
\label{good_dist_agreement}
\end{subfigure}
\begin{subfigure}[h]{0.94\linewidth}
\includegraphics[width=\linewidth]{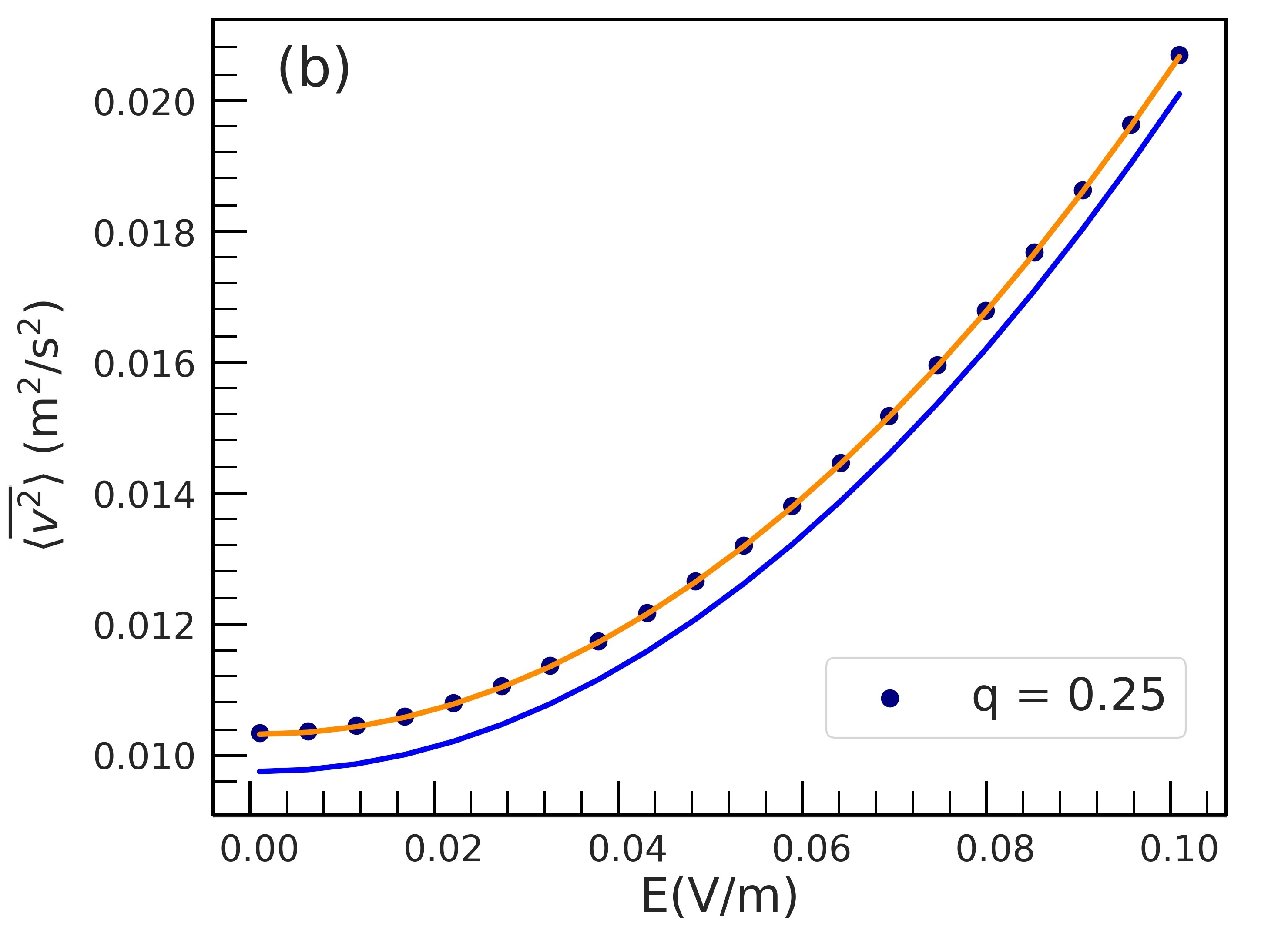}
\label{bad_dist_agreement}
\end{subfigure}%
 \caption{$\bar{\langle v^{2} \rangle}$ vs E. Panel (a) shows the dependence for two different values of the $q$-parameter, the points are the numerical results while the solid lines corresponds to approximate solution in Eq.(\ref{eq16}). Panel (b) focuses on the $q$=0.25 case, the solid lines are the fits using Eq.(\ref{eq16}) (blue) and Eq.(\ref{eq17}) (orange).  The rest of trap parameters are the same as those in Fig. (\ref{fig2}) for both figures }
\label{fig3}
\end{figure}

In the case of a constant excess micromotion Eq.~(\ref{correlations_Eq}) is solved numerically via a fourth-order Runge-Kutta  method and the results are shown in Fig.~\ref{fig2}, which depicts $\overline{\langle v^{2} \rangle}$ as a function of $q_{i}$, keeping $a$ and $\Omega$ fixed. This figure indicates that a constant excess micromotion distorts the ion's dynamics at small values of $q$. On the contrary, an excess of micromotion does not alter the ion's dynamics for large values of $q$. Therefore, a coupling must exist between $q$ and the excess micromotion. In particular, Eqs.~(\ref{time_avg_msr_1}) and (\ref{time_avg_msr_1_bis}) transform into

\begin{equation}
    \begin{split}
        &\overline{\langle x^{2} \rangle} = \frac{2|Q_{1}|^{2}\big(\frac{D}{m}+ 2\frac{\tilde{E}_{x}^{2}}{\omega^{2}}\gamma^{\prime}\big)-\frac{\tilde{E}_{x}^{2}q}{\omega^{2}}\text{Re}(Q_{0}^{+}Q_{-1}(1+i))}{Q_{0}|Q_{1}|^{2} - \text{Re}(Q_{0}^{+}Q_{1}^{-}Q_{-1}) } \\
        & \;\;\;\;\;\;\; \approx \frac{k_{\text{B}}T}{m\omega^{2}} +  \frac{\tilde{E}_{x}^{2}}{\omega^{4}} \\
        & \overline{\langle v^{2} \rangle}\approx \frac{2k_{\text{B}}T}{m} + \frac{\tilde{E}_{x}^{2}}{\omega^{2}},
    \end{split}
    \label{eq16}
\end{equation}

\noindent
establishing a quadratic dependence of the mean squared velocity of the ion on the applied electric field, as it is corroborated in Fig.~\ref{fig3}. This figure displays the dependence of $\overline{{\langle v^{2} \rangle}}$ concerning the constant external field, comparing numerical results versus the approximation in Eq.~(\ref{eq16}). As a result, and as expected, we notice that Eq.~(\ref{eq16}) fails to describe the average square velocity for $q$=0.25 adequately. Therefore, for a more accurate description of the ion's dynamics, it is necessary to go to the next level of approximation, yielding 

\begin{equation}
    \begin{split}
         &\overline{\langle x^{2} \rangle} \approx  \bigg( \frac{k_{\text{B}}T}{m\omega^{2}} + \frac{\tilde{E}_{x}^{2}}{\omega^{4}}\bigg)\bigg[ \frac{1+\frac{27}{50}q^{2}}{1-\frac{1}{5
         }q^{2}}\bigg]\\
         &\overline{\langle v^{2} \rangle} \approx  \frac{\Omega^{2}}{4}\bigg( \frac{k_{\text{B}}T}{m\omega^{2}} + \frac{\tilde{E}_{x}^{2}}{\omega^{4}}\bigg)\bigg[ a\bigg( \frac{1+\frac{14}{25}q^{2}}{1-\frac{24}{31
         }q^{2}}\bigg) + q^{2}\bigg( \frac{1+\frac{20}{25}q^{2}}{1-\frac{7}{50
         }q^{2}}\bigg) \bigg] \\
         & \;\;\;\;\;\;\; -\frac{\tilde{E}_{x}^{2}}{\omega^{2}}\bigg( 1 + \frac{q^{2}}{2}\bigg),
    \end{split}
    \label{eq17}
\end{equation}
which show a similar $q$-dependence as in Eq.~(\ref{time_avg_msr_2}), although a new field-dependent term appears. Eq.~(\ref{eq17}) describes accurately the mean square velocity of the ion for $q$ values as large as 0.25, as it is shown in panel (b) of Fig.(\ref{fig3}).

Moreover, we notice that higher-order terms of the Fourier expansion for the time-averaged mean squared distance with and without excess micromotion are related (see table \ref{fourier_terms}) as

\begin{equation*}
    x_{n+1,E_{\text{mm}}} \approx x_{n+1} + \frac{x_{n+1}}{x_{0}}\frac{\tilde{E^{2}_{x}}}{\omega^{4}} + S_{n}^{+}\Big( 1 + \frac{2\gamma^{\prime}}{\Omega}\ii \Big)\frac{\tilde{E}^{2}_{x}q}{\Omega^{2}\omega^{2}}, 
\end{equation*}
where $x_{n+1}$ and $x_{n+1,E_{\text{mm}}}$ denote the Fourier coefficients in the absence and presence of $E_{\text{mm}}$, respectively, and $S^{+}$ is given by Eq.~(\ref{S_coefficients}). Therefore, the contribution of the $n$th-harmonic of the trap frequency depends not only on the $q$-parameter but also on the magnitude of the stray field causing the constant excess micromotion.

Finally, we study a more general scenario including a time-dependent excess micromotion: F$_{\text{mm}e,x}$ = $e(E_{\text{dc,}x} + E_{{\text{ac,}x}}$sin($\Omega$t)), where $E_{\text{dc}}$ is an stray field, as previously considered, and $E_{\text{ac}}$sin($\Omega$t) stands for the new time-dependent component, where the amplitude $E_{\text{ac}}$ depends on the trap parameters. The mean position of the ion is found by solving Eq.~(\ref{expected_white}) that leads to

\begin{equation}
\label{eq18}
\begin{split}
    \langle x \rangle \approx & \bigg(\frac{\tilde{E_{x}}}{\omega^{2}} - \frac{\tilde{E_{x}}}{\omega^{2}} \text{e}^{\frac{-\gamma^{\prime} t}{2}}\cos(\omega_{d}t) \bigg)\bigg(1+\frac{q_{x}}{2}\cos(\Omega t)\bigg) \\ & + \frac{\tilde{E}_{\text{ac}}}{\Omega^{2}}\sin(\Omega t),
\end{split}
\end{equation}
where $\tilde{E}_{\text{ac}}$=$eE_{\text{ac}}/m$; and the mean velocity is obtained upon derivation of Eq.~({\ref{eq18}}) as a function of time. Consequently, the time average values of $\langle x^{2} \rangle$ and $\langle v^{2} \rangle$ read as

\begin{equation}\label{ms_with_ac_field}
    \begin{split}
        &x_{0} \approx \frac{k_{\text{B}}T}{m\omega^{2}} +  \frac{\tilde{E}_{\text{dc}}^{2}}{\omega^{4}} + \frac{\tilde{E}_{\text{ac}}^{2}}{2\Omega^{2}\omega^{2}} - \frac{\tilde{E}_{\text{dc}}\tilde{E}_{\text{ac}}q\Omega}{16\omega^{4}\gamma^{\prime}}  \\
        & v_{0}\approx \frac{2k_{\text{B}}T}{m} + \frac{\tilde{E}_{\text{dc}}^{2}}{\omega^{2}} + \frac{\tilde{E}_{\text{ac}}^{2}}{\Omega^{2}} - \frac{\tilde{E}_{\text{dc}}\tilde{E}_{\text{ac}}q\Omega}{8\omega^{2}\gamma^{\prime}},
    \end{split}
\end{equation}

\noindent
which are different from the case of a constant electric field. In Eq. (\ref{ms_with_ac_field}), it is worth emphasizing that the last term coupling the time-dependent component with the time-independent one of stray fields with the Paul trap fields is usually ignored in other derivations~\cite{Berkeland}. 

 \begin{figure*}[t]
       \centering
\begin{subfigure}[b]{0.36\textwidth}
\includegraphics[width=\textwidth]{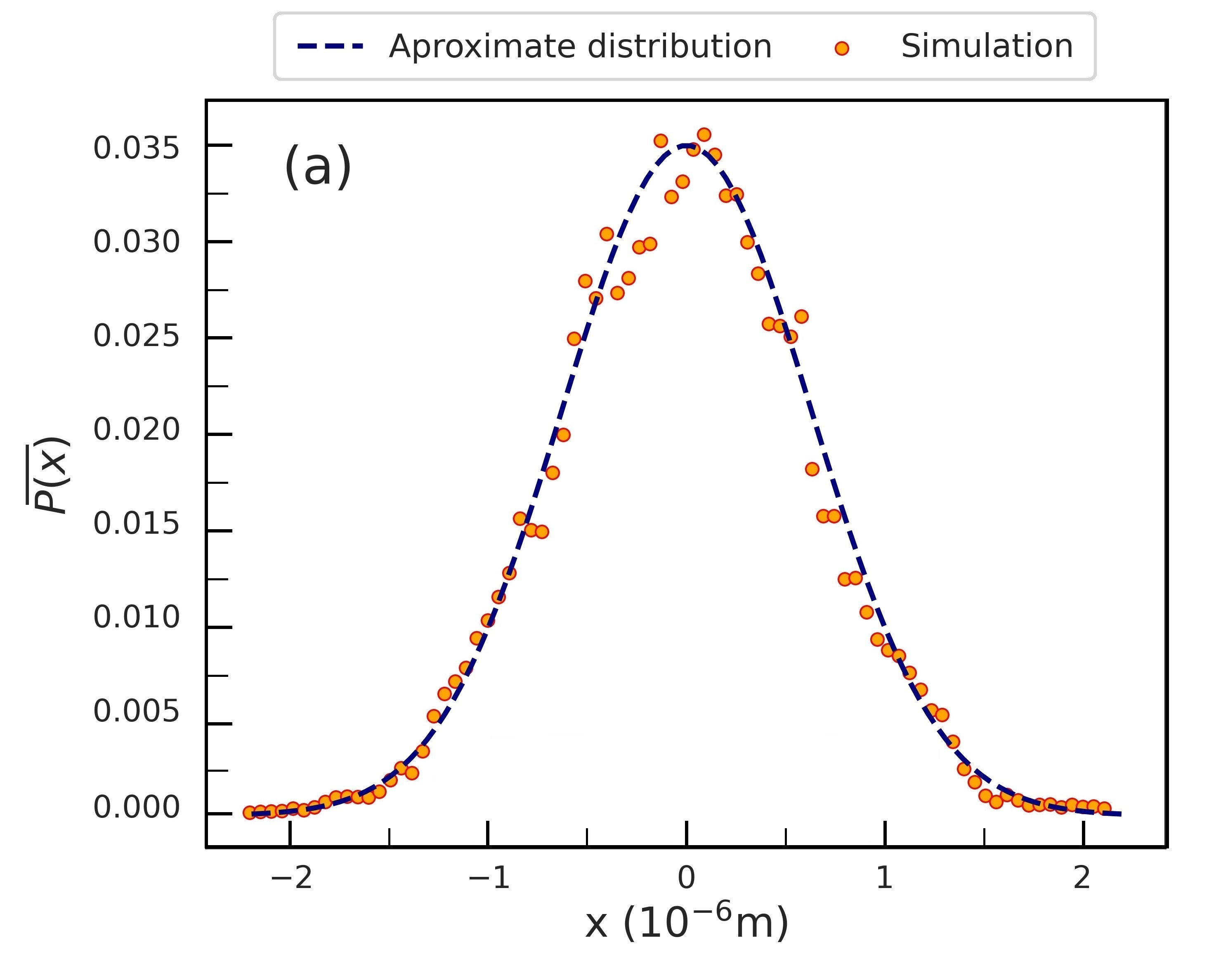}
\label{spatial_distribution_no_field}
\end{subfigure}
\begin{subfigure}[b]{0.36\textwidth}
\includegraphics[width=\textwidth]{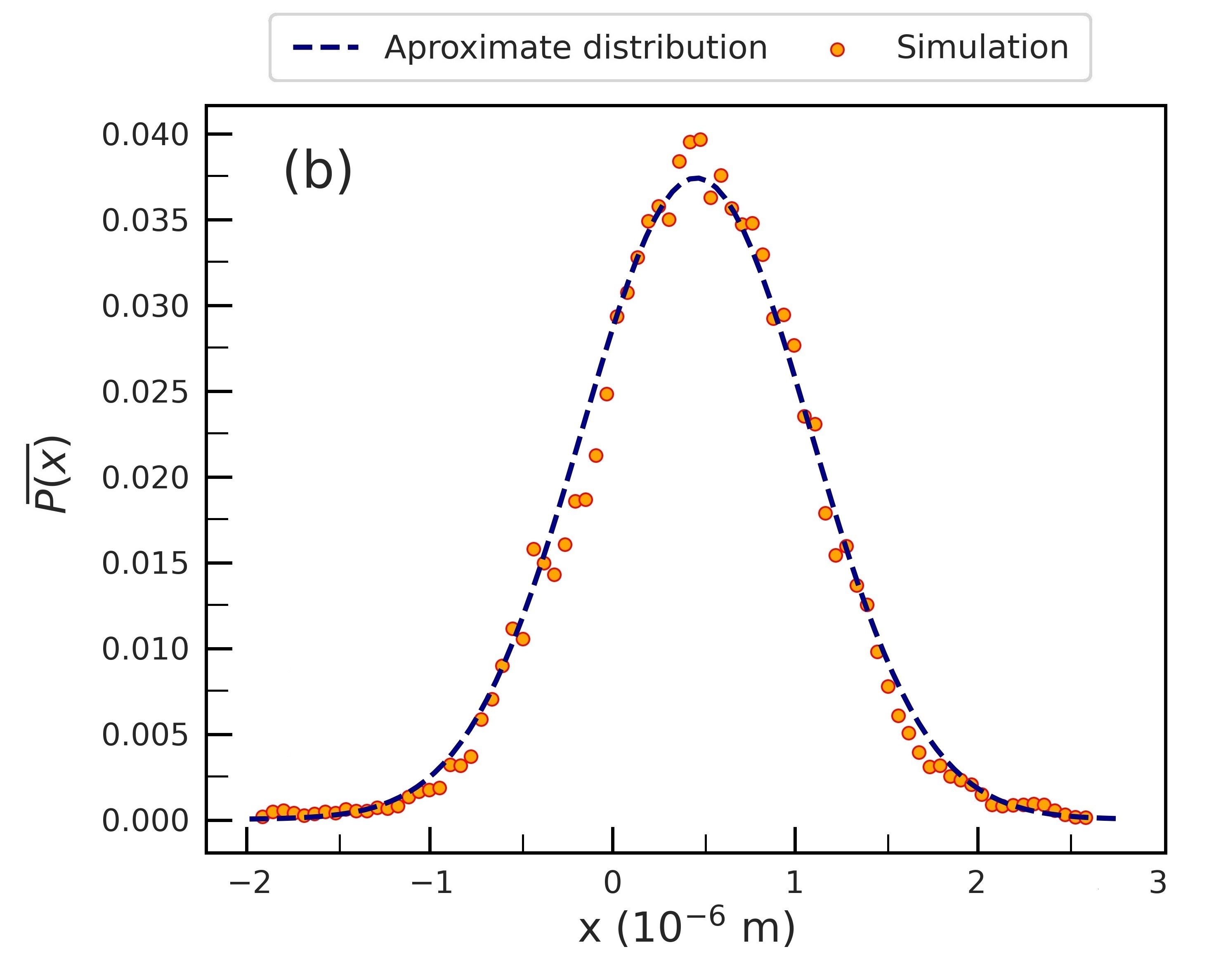}
\label{spatial_distribution_field}
\end{subfigure}
\begin{subfigure}[b]{0.36\textwidth}
\includegraphics[width=\textwidth]{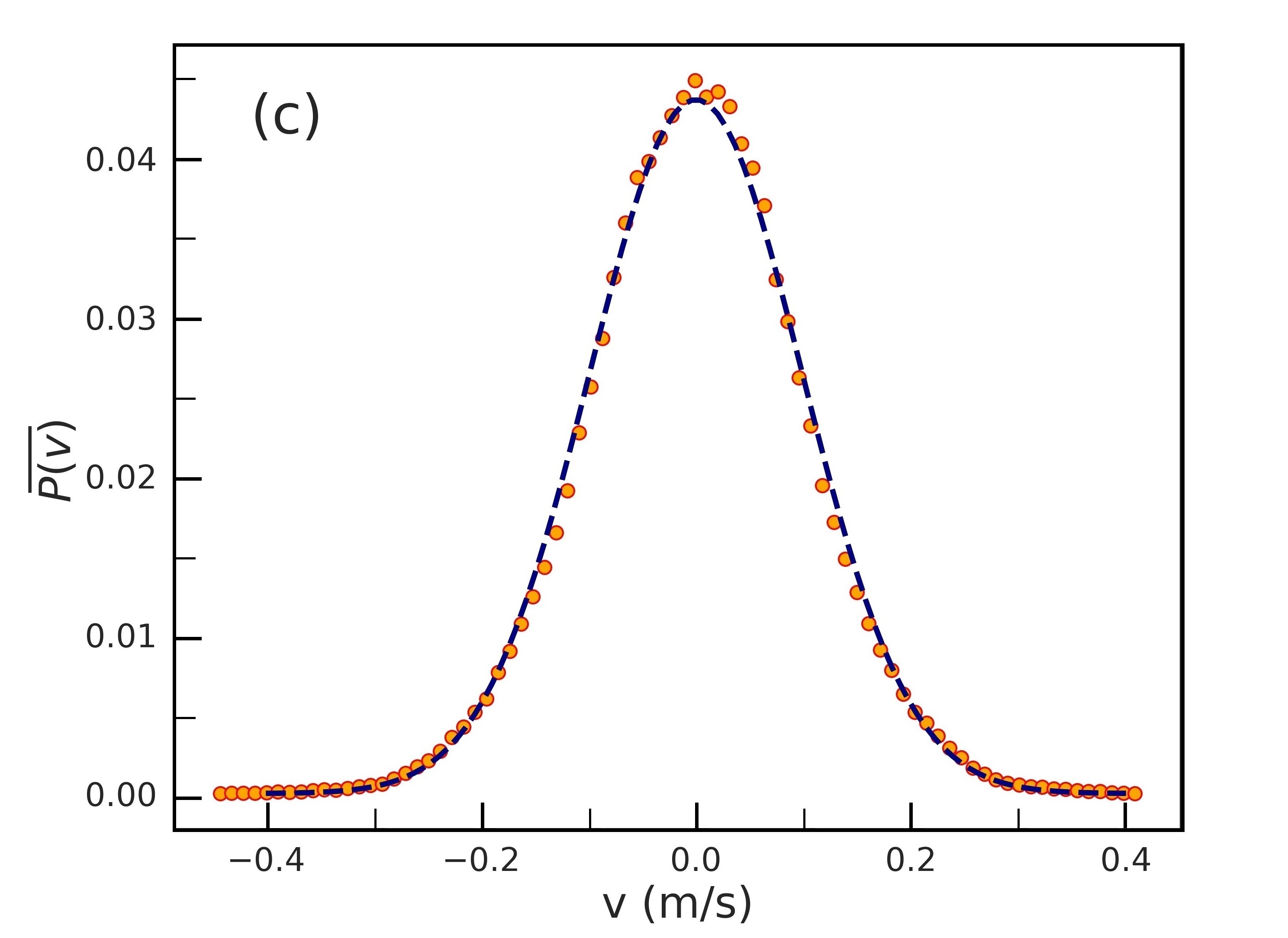}
\label{velocity_distribution_no_field}
\end{subfigure}
\begin{subfigure}[b]{0.36\textwidth}
\includegraphics[width=\textwidth]{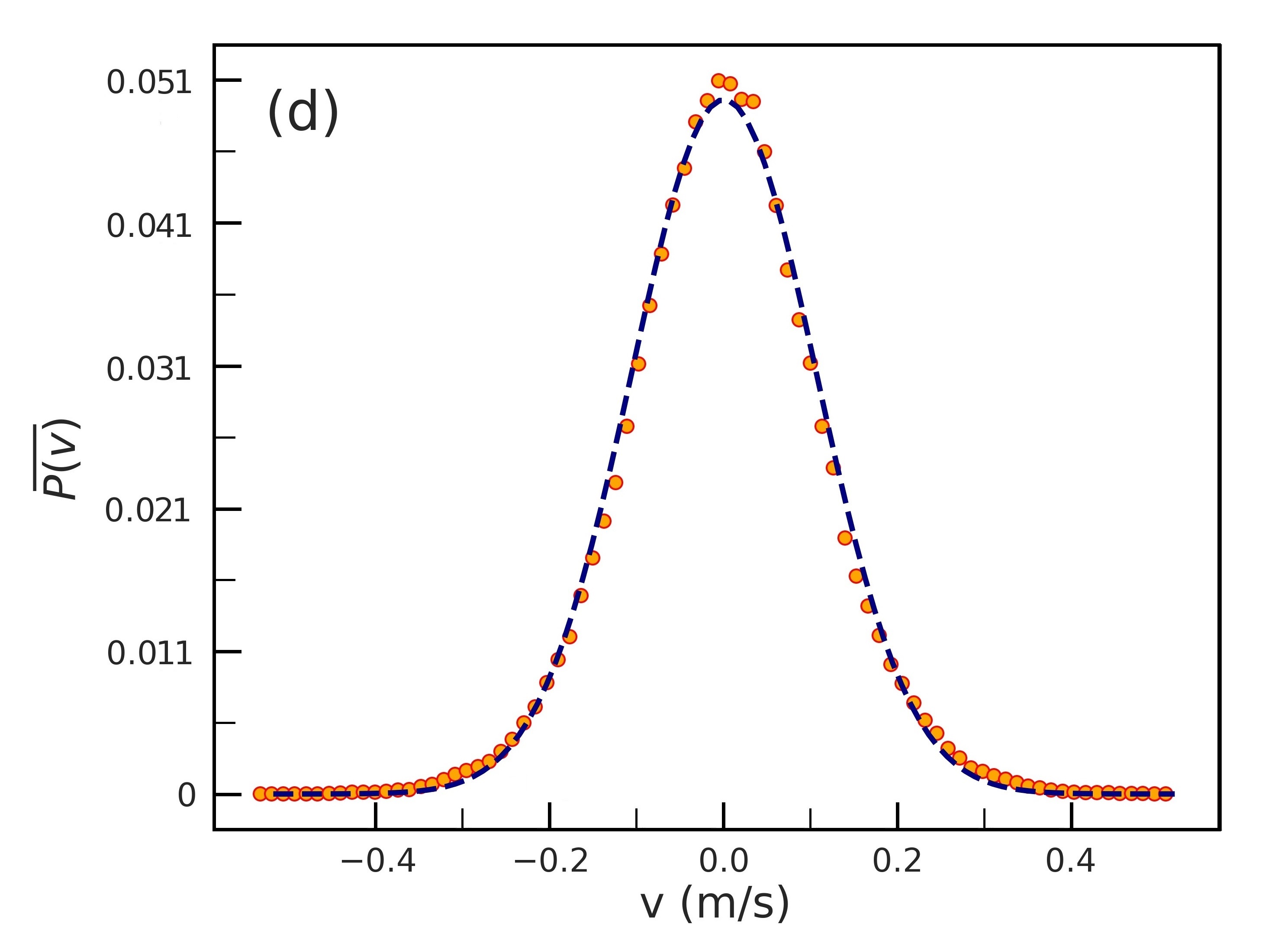}
\label{velocity_distribution_field}
\end{subfigure}
\caption{Probability density function for ion's position and velocity after solving the Langevin equation. The fitting distribution are the approximate time average distributions previously derived in Eqs.~(\ref{AD_field}) and (\ref{VAD_field}). Panels (a) and (b) present the $x$-position distribution without a stray field and with a stray field of 0.01 V/m respectively. Panels (c) and (d) present the $v$-distribution for the same stray field conditions as in (a) and (b) respectively. For all the simulations we use the trap parameters $a=$-8$\times10^{-6}$, $q=0.1$ and $\Omega = 2\pi\times10^{6}\,$Hz,a total of $1\times10^{6}$ points were used for the distributions. Again, a  $^{171}$Yb$^{+}$ -$^{6}$Li mixture was considered in the calculations.} 
\label{fig4}
    \end{figure*}

\subsection{Distributions}\label{distributions_sec}
As a consequence of the time-dependent trapping potential on the ion, the system reaches a quasi-stationary state rather than a stationary state, in which the trap frequency drives the time evolution. The phase space distribution for each component in the long time limit is given by

\begin{equation*}
\begin{split}
    P(x,v,t) = & \frac{1}{\sqrt{(2\pi\sigma_{xx})(2\pi\sigma_{vv})(1 - \rho_{12}^{2})}}\times \\
    & \text{exp}\bigg[ -\frac{1}{1-\rho_{12}^{2}} \bigg( \frac{\delta x ^{2}}{2\sigma_{xx}} - \frac{\rho_{12}\delta x \delta v}{\sqrt{\sigma_{xx}\sigma_{vv}}}   + \frac{\delta v ^{2}}{2\sigma_{vv}} \bigg)  \bigg]
\end{split}
\end{equation*}
where $\sigma_{xx}$ ($\sigma_{vv}$) represents the variance of the position (velocity), $\delta x$ ($\delta v$) are the displacement from the mean value $\langle x \rangle$($\langle v \rangle$) and $\rho_{12}$ is the normalized correlation $\frac{\sigma_{xv}}{\sqrt{\sigma_{xx}\sigma_{vv}}}$; all of these parameters are time-dependent. Therefore, the time-averaged position distribution reads as

\begin{equation}\label{TAD}
\begin{split}
\overline{P(x)} &= \frac{\Omega}{2\pi}\int_{0}^{\frac{2\pi}{\Omega
}}\int_{-\infty}^{\infty} P(x,v,t) dv dt \\ &
= \frac{\Omega}{2\pi}\int_{0}^{\frac{2\pi}{\Omega
}}\frac{dt}{\sqrt{2\pi\sigma_{xx}}}\text{exp}\bigg[ -\frac{\delta x ^{2}}{2(1-\rho_{12}^{2})\sigma_{xx}} \bigg( 1 - \frac{\sigma_{xv}^{2}}{\sigma_{vv}} \bigg)\bigg],
\end{split} 
\end{equation}
which, unfortunately, can not be solved analytically. However, an approximate time averaged distribution can be derived using the average mean square position and velocity from Eqs.~(\ref{eq16}) and (\ref{ms_with_ac_field}) for the case of time-independent and time-dependent excess of micromotion, respectively. Then, the approximate time averaged distribution for $x$ is
\begin{equation}\label{AD_field}
        \overline{P(x)} \approx \sqrt{\frac{m \omega^{2}}{2\pi k_{\text{B}}T}}\exp \Bigg[\frac{-m\omega^2 \Big( x - \frac{\tilde{E_{x}}}{\omega^{2}} \Big)^{2} }{2k_{\text{B}}T}\Bigg],
\end{equation}

\noindent
whereas, the approximate velocity distribution becomes 
\begin{equation}\label{VAD_field}
    \overline{P(v)} \approx \sqrt{\frac{1}{{2\pi \bigg( \frac{2k_{\text{B}}T}{m} + \frac{\tilde{E_{x}}^{2}}{\omega^{2}}\bigg)}}}\exp \Bigg[-\frac{1}{2}\frac{ v_{x} ^{2} }{\Big( \frac{2k_{\text{B}}T}{m} + \frac{\tilde{E_{x}}^{2}}{\omega^{2}}\Big)}\Bigg]. 
\end{equation}

The accuracy of Eqs.~(\ref{TAD}) and (\ref{VAD_field}) is contrasted in Fig.~\ref{fig4}, in which the numerical results for the probability distribution for the ion position and velocity are shown. As a result, it is observed that the derived approximated expressions perform exceptionally well compared to numerical results.

\section{Colored noise bath}
\label{colored_noise_sec}

In the case of a bath with a memory effect, the ion's dynamics is described by the GLE [ Eq.(\ref{GLE})]. Hence, the probability distribution $P(x,v,t)$ is no longer Markovian, and its statistical properties depend on the type of noise chosen to characterize the bath. In particular, we consider a Gaussian noise with an exponentially decaying correlation function of the form
\begin{equation}
\label{correlation_func}
    \langle \zeta_{\text{c}}(t)\zeta_{\text{c}}(s) \rangle = \frac{D}{\tau_{\text{c}}}\exp \bigg(-\frac{|t-s|}{\tau_{\text{c}}}\bigg),
\end{equation}
where the strength $D$ and the correlation time ($\tau_{\text{c}}$) depend on the atom-ion scattering properties. This noise is described as an Ornstein-Uhlebeck (OU) process as 
\begin{equation}\label{OU}
\dot{\zeta_{\text{c}}} = -\frac{1}{\tau_{\text{c}}}\zeta_{\text{c}} + \frac{\sqrt{D}}{\tau_{\text{c}}}\zeta,   
\end{equation}
where $\zeta$ represents a {\it white noise}. Then, to elucidate the ion's dynamics, one needs to solve a set of integro-differential stochastic equations constituted by Eq.~(\ref{GLE}), including the noise effects through Eq.~(\ref{OU}). 

\subsection{Correlation time}

The correlation time is estimated as the atom-ion collisional time $\tau_{\text{c}} = \frac{1}{\Gamma_{\text{L}}}$, in which

\begin{equation*}
    \Gamma_{\text{L}} = \rho k_{\text{L}},
\end{equation*}
is the Langevin rate for atom-ion collisions, $\rho$ is the density of the bath and $k{_\text{L}}$ is the Langevin reaction rate given by~\cite{J_perez_book}

\begin{equation*}
    k_{\text{L}} = 2\pi \sqrt{\frac{C_{4}}{\mu}},
\end{equation*}
where $\mu$ is the atom-ion reduced mass and $C_4=\alpha/2$ (in atomic units), being $\alpha$ the polarizability of the atoms from the bath. It is worth noticing that even though the Langevin model for charged-neutral collisions is based on a classical framework, it applies to temperatures as low as $100 \, \mu K$ (depending on the mass of the colliding partners)\cite{R.Cote}.

\subsection{Ion dynamics}

The mean position and velocity are determined in a two step approach: first Eq.~(\ref{OU}) is solved using the stochastic method explained in Appendix~\ref{appendixB}, generating a $colored \,\, noise$ vector ($\vec{\zeta_{\text{c}}}$). Second, the noise vector is used in the GLE, which is solved using a finite difference method (see Appendix~\ref{appendixB}). In particular, in each run we generate $10^{6}$ realizations to generate average evaluations for any dynamical observable.

\begin{figure}[]
\centering
\hspace{-0.1cm}
\includegraphics[scale=0.12
]{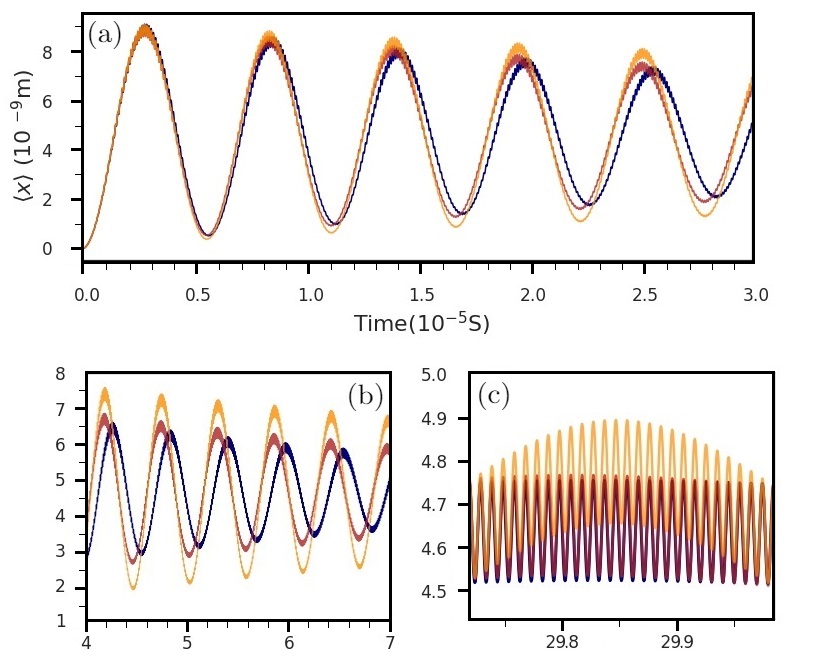}
\caption{ Expected position for the $^{171}$Yb$^{+}$ ion in the $^{6}$Li cloud using white noise (blue) and colored noise with correlation times equal to $\tau$ = 1.04$\times$ 10$^{-6}\,$s (red) and  $\tau$ = 1.54$\times$ 10$^{-6}\,$s (yellow). All the trap parameters are equal for each simulation: $q$ = 0.1, $a$= -8$\times10^{-6}$, $\Omega$ = 2$\pi\times10^{-6}\,$Hz and E$_{\rm{mm}}$ = 0.01 V/m.}  
\label{Expected_x_white_and_Color}
\end{figure}

Fig.~\ref{Expected_x_white_and_Color} displays the evolution of the ion's mean position considering white and colored noises (with two different correlation times). The addition of a new time scale associated with $\tau_{c}$ introduces a delay in the relaxation time of the processes, hindering the evolution into a quasi-steady state. Similarly, the colored noise results slightly increase the secular frequency, which can be seen as a retarded effect due to the memory kernel. Moreover, the colored noise induces oscillations in the observables' evolution, even in the case of a simple free Brownian particle \cite{Schmidt2015}: $memory \,\, oscillations$ are coupled to the secular ones, resulting in a modified secular displacement. 

\begin{figure}[]
\centering
\hspace{-0.1cm}
\includegraphics[scale=0.12
]{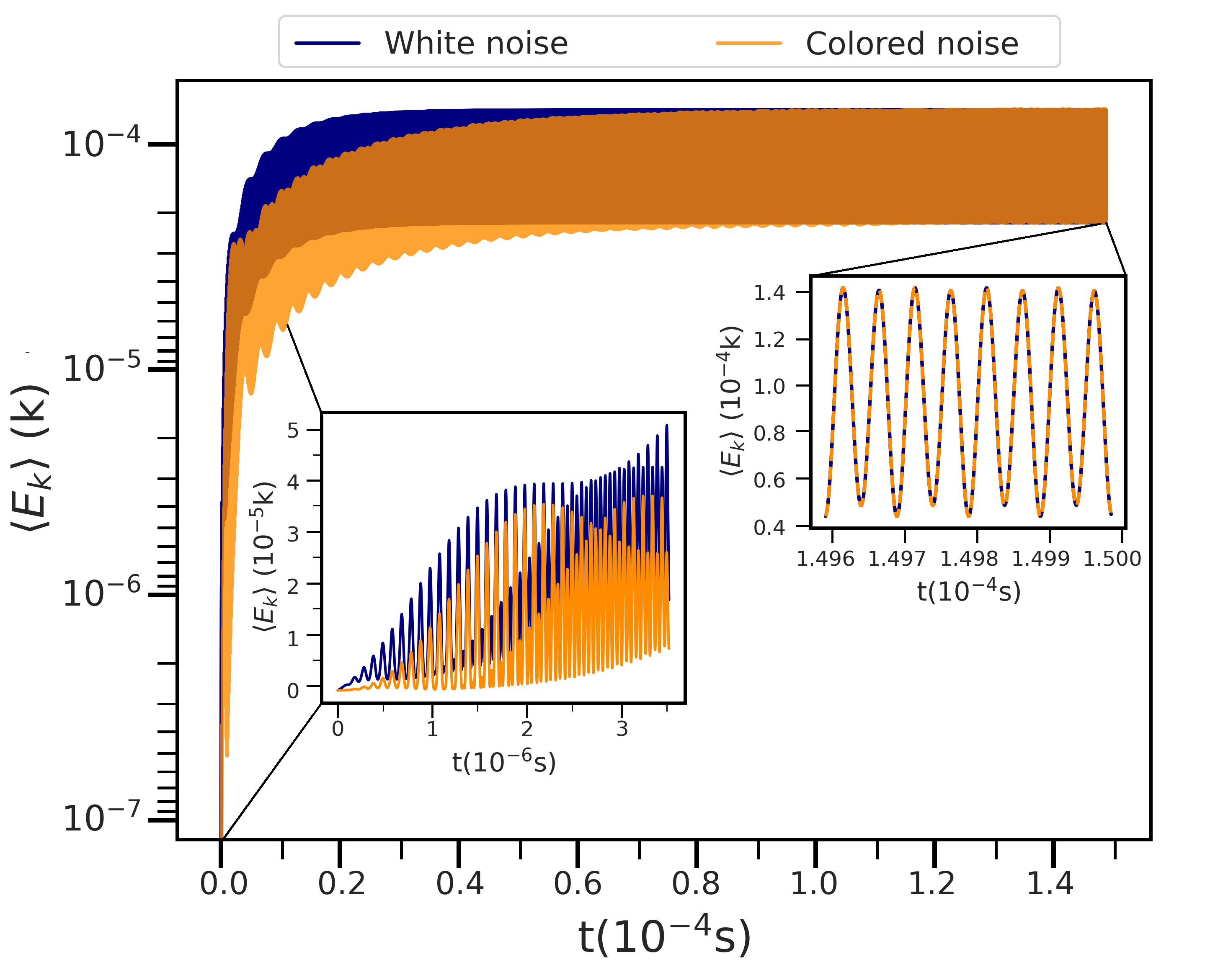}
\caption{Energy relaxation for two different noise models in a $^{6}$Li -$^{171}$Yb$^{+}$ system. The agreement of both descriptions, for a given atom-ion mixture, depends of the atomic density}
\label{Energy_ralaxation_comparison}
\end{figure}

 Fig.~\ref{Energy_ralaxation_comparison} presents the relaxation process of the kinetic energy for white and color noises. As a result, it can be noticed that the quasi-stationary behavior of the two formulations coincides. Thus, the average mean values for colored and white noise are equivalent. Similarly, since the GLE is linear in $x$, $v$, and the noise, the phase space distribution $P(x,v,t)$ will show a similar shape as the one described for the white noise case in Fig.~\ref{fig4}. On the other hand, the time scales associated with the friction coefficient and the bath correlations depend on the atomic density. Therefore, by varying the density of the atomic gas, it is possible to tailor the role of memory effects. In particular, for a given ion, correlation effects increase with the mass-to-polarization ratio of the atom.

\section{\label{molecular_dynamics_sec}Monte Carlo Simulations}

In addition to the Langevin equation formulation, we have also implemented a Monte Carlo simulation of the ion-atom dynamics following Ref. \cite{Zipkes}. This simulation works only in the regimen of pseudopotential approximation and assumes that atom-ion collisional cross section are well described by the Langevin one. We compare the results obtained with this formulation and the stochastic dynamics.

\begin{figure}
\begin{subfigure}[h]{1.08\linewidth}
\includegraphics[width=0.9\linewidth]{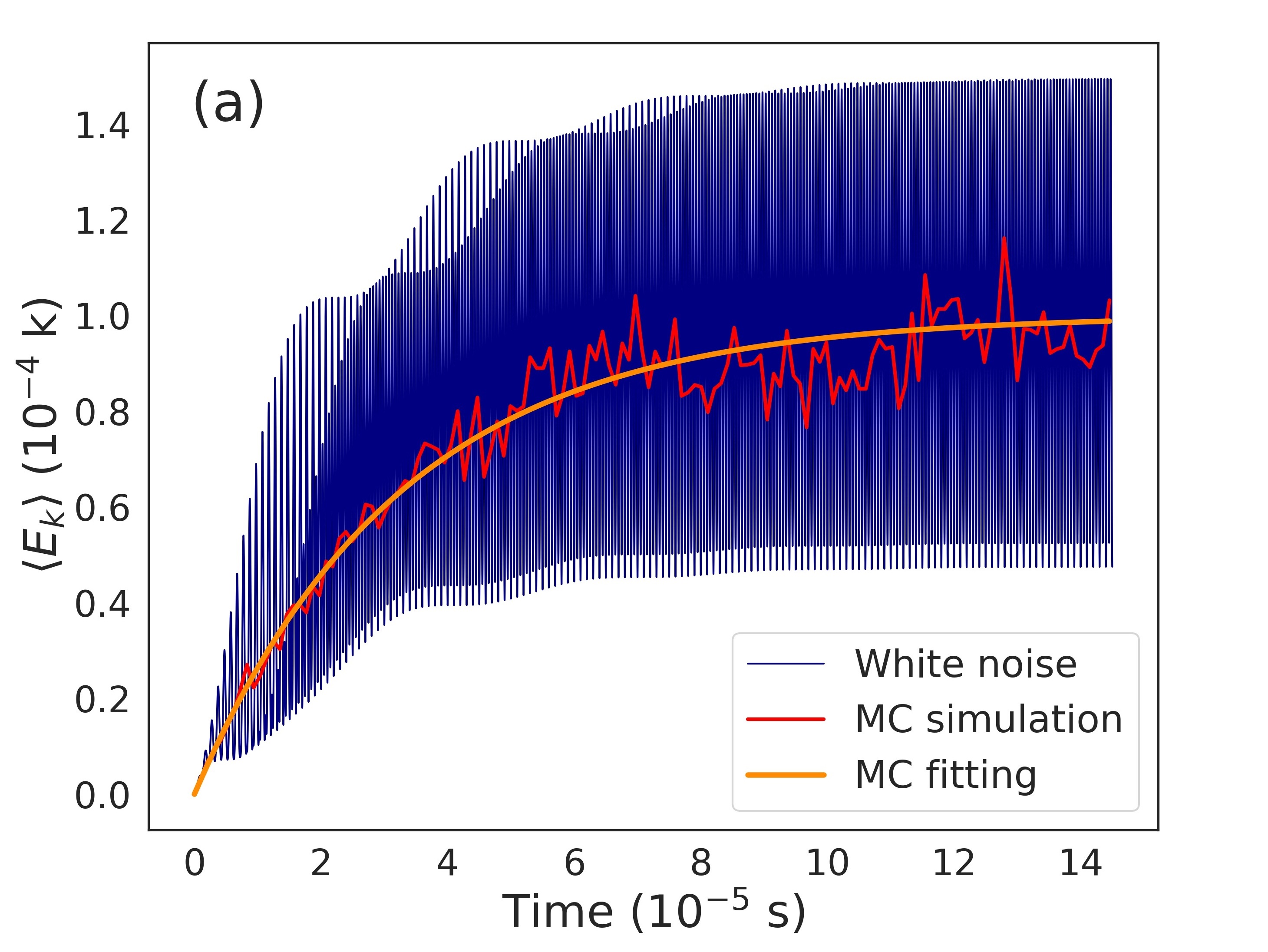}
\label{sqv_vs_q_no_field}
\end{subfigure}
\begin{subfigure}[h]{0.98\linewidth}
\includegraphics[width=\linewidth]{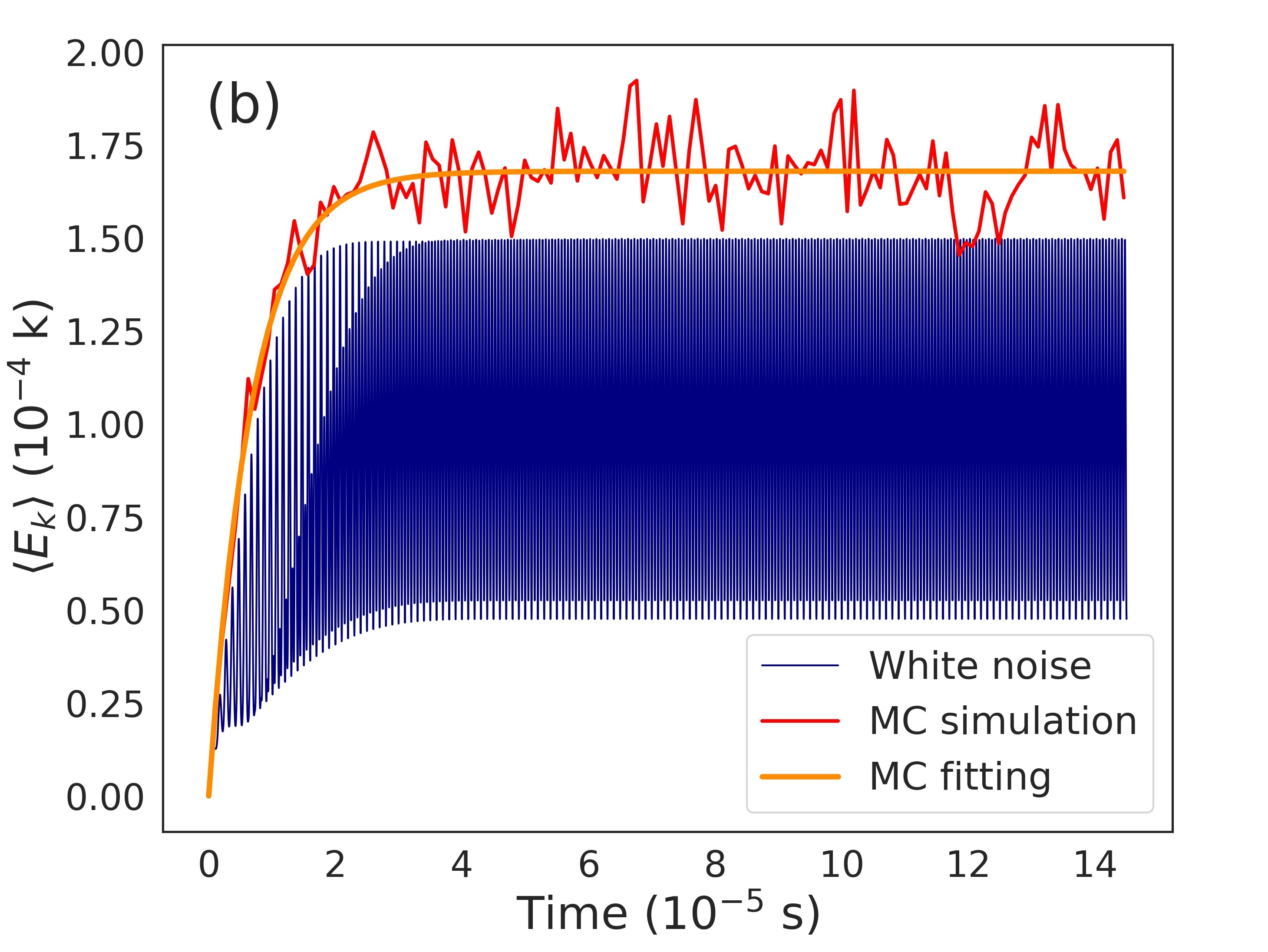}
\end{subfigure}%
\caption {Average $x$-contribution kinetic energy  as a function of time of a single $^{171}$Yb$^{+}$ ion immersed in an atomic bath based on MC simulations and on solving the LE with white noise: (a) $^{6}$Li and (b) $^{87}$Rb }
\label{Energy_relaxation_LD_vs_MD}
\end{figure}

The time evolution of $x$-component contribution to the kinetic energy for both white noise Langevin (LE) and Monte Carlo (MC) simulations for two different atomic baths are displayed in Fig. \ref{Energy_relaxation_LD_vs_MD}. As a result, we notice that the LE simulations show the oscillatory behavior in the energy evolution due to the trapping potential, which has to do with the intrinsic continuous-time nature of the Langevin equation. Nevertheless, the average value of Langevin predictions agrees with MC simulations in the case of an atomic bath of $^{6}$Li. However, there is a discrepancy in the case of $^{87}$Rb. Indeed, Langevin's model predictions look almost the same independently of the mass of the atoms in the bath since only the ion mass explicitly appears in the Langevin equation. On the contrary, in MC simulations, the relevant parameter is the atom-ion reduced mass. We extract the relaxation times by fitting the MC results to a time-dependent decaying exponential for a more exhaustive comparison between both simulation approaches. For the $^{6}$Li atomic bath, we obtained relaxations times of $23.00\,\mu $s and $22.79\,\mu $s for the Langevin and Monte Carlo simulations, respectively, resulting in a relative error of 9.36\%  between the two schemes. In contrast, the error associated with the average kinetic value for long times is just 0.3\%. However, for a $^{87}$Rb atomic bath, the relaxation times and the average kinetic energy becomes very different between both formulations; the associated error for the mean kinetic energy is more significant than 50\% and even more prominent for the relaxation times, resulting in a value of $8.16\, \mu s$ for the Langevin simulation and less than  $4\, \mu s$ for the Monte Carlo scheme.

\begin{figure}
\begin{subfigure}[h]{1.08\linewidth}
\includegraphics[width=0.9\linewidth]{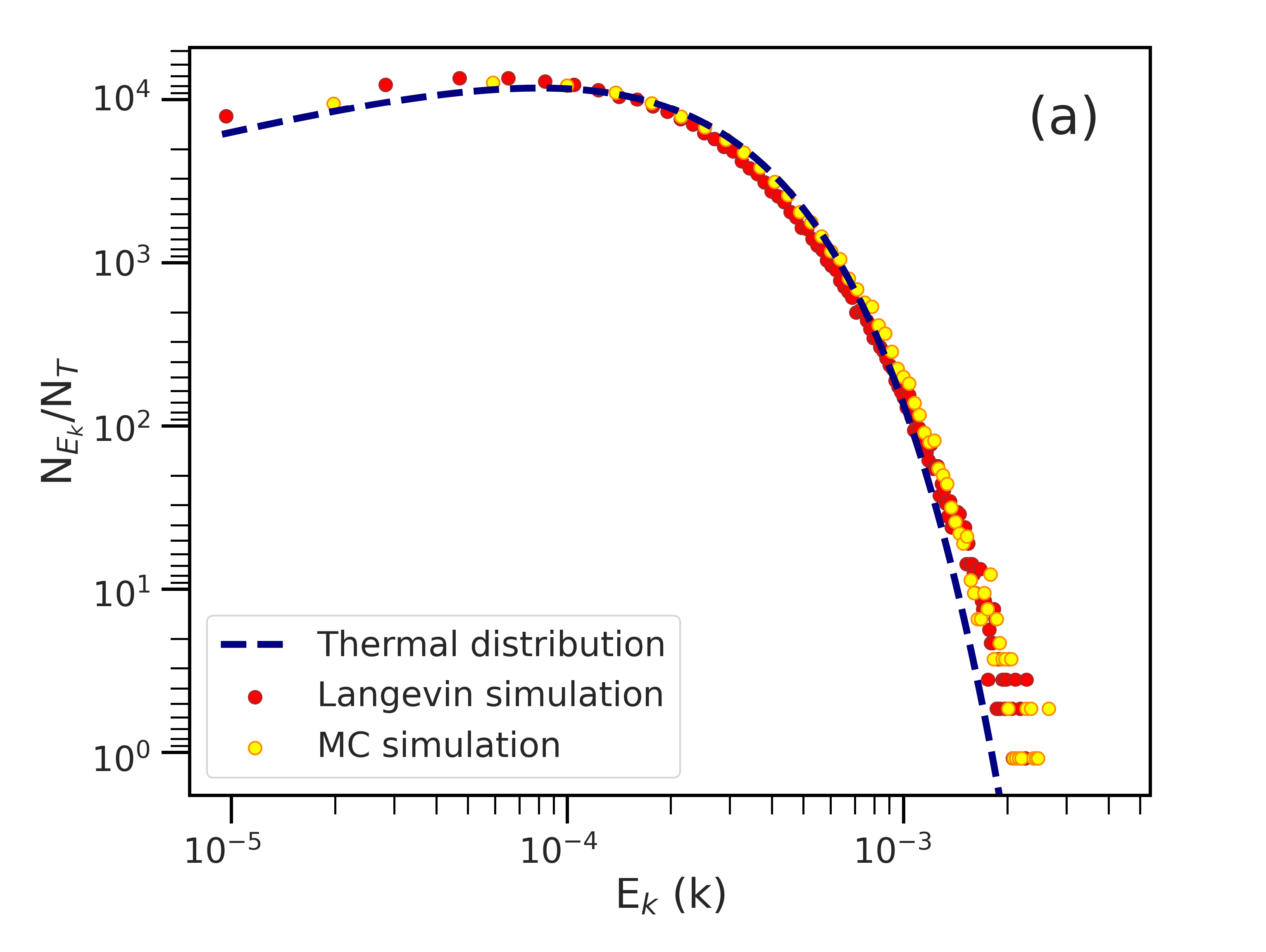}
\label{sqv_vs_q_no_field}
\end{subfigure}
\begin{subfigure}[h]{0.98\linewidth}
\includegraphics[width=\linewidth]{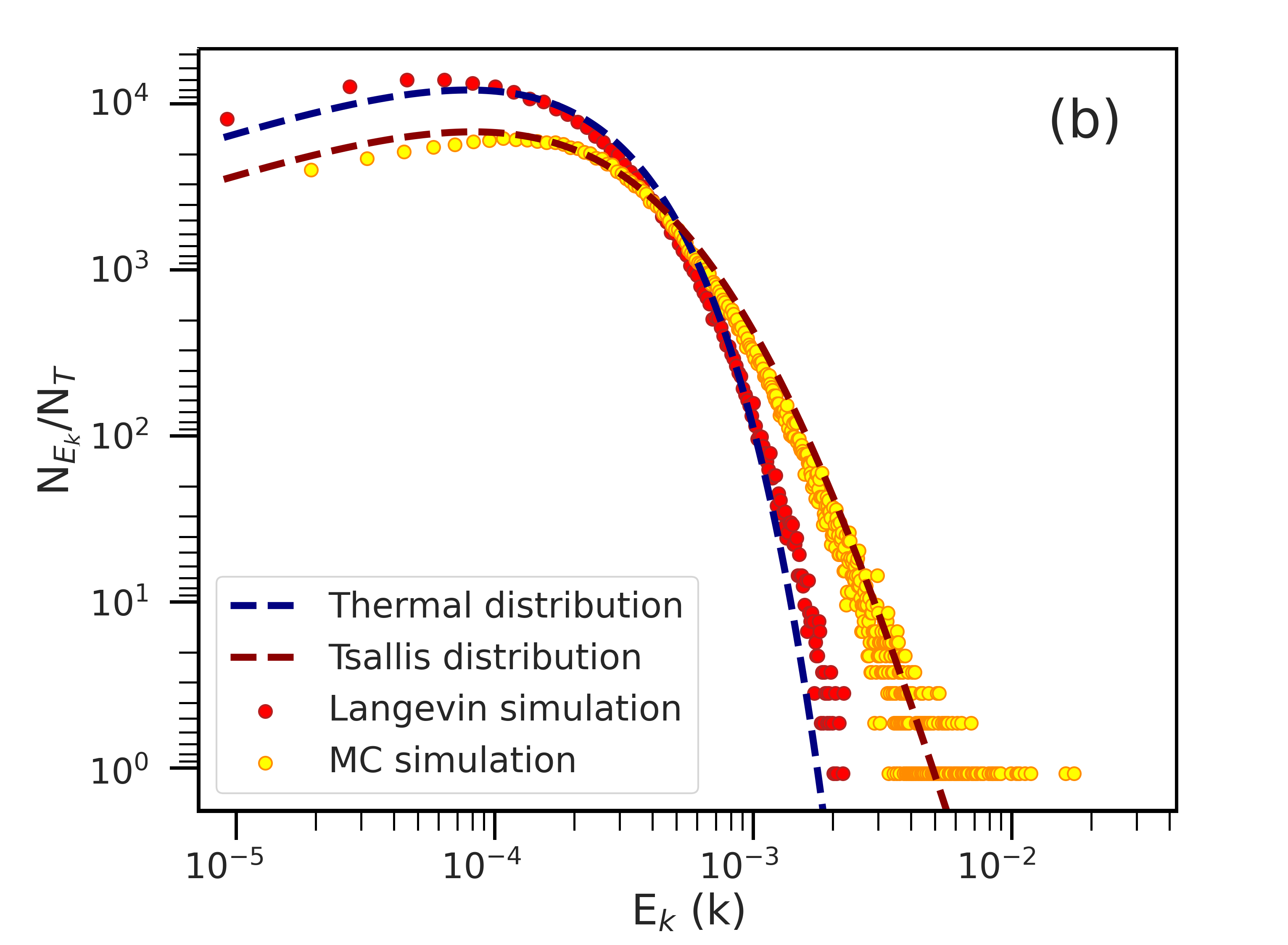}
\end{subfigure}%
\caption {Kinetic energy distribution of the $^{171}$Yb$^{+}$ ion in an atomic cloud of (a) $^{6}$Li and (b) $^{87}$Rb, both at T = 10$^{-4}\, \text{K}$ }
\label{KE_distributions}
\end{figure}

Fig.~\ref{KE_distributions} present the kinetic energy distribution of an $^{171}$Yb$^{+}$ ion immersed in two different baths. For the $^{6}$Li bath, panel (a), both formulations lead to a thermal distribution characterized for the atomic cloud temperature ($T_{\text{a}}$), with an ion kinetic temperature of $T_{\text{kin}} \approx 3T_{\text{a}}/5$. On the other hand, MC simulations for the $^{87}$Rb bath  (panel (b) of Fig.~\ref{KE_distributions}) reveal a different behavior: the kinetic energy distribution follows a Tsallis distribution of the form \cite{Pinkas2020}

\begin{equation}
   P(E_{k}) = A  \frac{(n-2)(n-3)(n-4)}{2(nk_{\rm{B}}T)^{3}}\frac{E_{k}^{1/2}}{\big(1+\frac{E_{k}}{nk_{\rm{B}}T}\big)^{n}},
\end{equation}
where $A$, $n$ and $T$ are fitting parameters that for the case at hand are  1.2$\times$10$^{-4}$, 6.0 and 7$\times$10$^{-4}\,\text{K}$ , respectively. The parameter $T$ represents the physical temperature in the limit of the thermal distribution ($n\rightarrow \infty$). This power-law behavior is a result of the so-called $micromotion \,\, heating$ and depends strongly on the atom-to-ion mass ratio and the trap parameters \cite{Pinkas2020}. However, the stochastic formulation still leads to a thermal distribution due to the noise statistics and the additive nature of the stochastic equation of motion. Therefore, these results indicate that stochastic formulations for describing the ion dynamics in a bath are only applicable in the regimen of a low atom-to-ion mass ratio at which the heating effects are almost negligible but not null, as observed in Fig.~\ref{KE_distributions}.

\section{\label{conclusions}Conclusions}

We have studied the dynamics of a single trapped ion in a bath of ultracold atoms via the Langevin equation. The bath induces a stochastic force whose amplitude is controlled by atom-ion scattering properties. In addition, we have obtained analytical solutions for the average mean squared position and kinetic energy of the ion up to values $q \leq 0.3$. Thus, going beyond that any previous attempt, to the best of our knowledge. Similarly, the present formulation allows us to study the impact of excess micromotion (time-dependent and time-independent ), leading to analytical expressions for relevant magnitudes. In particular, we find that the excess micromotion adds energy to the system because of the time-varying trap potential. Moreover, we have derived an approximated position and velocity distribution for the ion with excess micromotion within the pseudopotential regimen. As a result, it is possible to estimate any necessary time-dependent or average property of the system.

The effects of the bath nature on the ion dynamics have been studied, comparing results from a white noise bath versus a colored noise bath. In particular, we have noticed that a colored noise bath requires solving the generalized Langevin equation. Therefore, it entails more physical information on the atom-ion few-body physics. In particular, the correlation time of the colored noise plays an essential role in the relaxation time scale of the ion. Although, those effects are mitigated by choosing an atomic species with a low mass-to-polarization ratio.

Finally, the results of the stochastic formulation have been tested against Monte Carlo simulations. The two formulations describe similar energy evolution and distributions in the low atom-to-ion mass ratio regimen, where the $micromotion \,\, heating$ reduces its contribution. The stochastic formulation turns out to be computationally cheaper than Monte Carlo simulations. For instance, $10^{5}$ collision events for $^{6}$Li-$^{171}$Yb$^{+}$, with n = $2\times10^{20}\,$m$^{-3}$ takes a CPU time of $461.66\,$s, whereas it takes $331.46\,$s for the Langevin simulation. This difference becomes clearer with the increasing number of collisional events. Additionally, it offers the possibility to explore the spectral aspects of the time evolution or the consideration of any bath correlation, which can be associated with the physical characteristics of the bath.

\appendix

\section{ Ion-atom few-body physics}
\label{appendixA}

The diffusion cross section is given by~\cite{Mott}
\begin{equation}
\label{eqdiff}
    \sigma_{\rm{D}}(E_k)=\int \frac{\text{d}\sigma_{\textrm{el}}(E_k)}{\text{d}\Omega}(1-\cos{\theta})\text{d}\Omega,
\end{equation}
where $\frac{\text{d}\sigma_\textrm{el}(E_k)}{\text{d}\Omega}$ represents the elastic differential cross section either classical or quantal, $\text{d}\Omega=2\pi \sin \theta \text{d} \theta$ the solid angle element, and $\theta$ is the scattering angle. Assuming a quantum mechanical description of the scattering observables, Eq.(\ref{eqdiff}) can be written as~\cite{Mott}
\begin{equation}
\label{sig_D}
    \sigma_{\text{D}}(E_k)=\frac{4\pi}{k^2}\sum_{l=0}^\infty(l+1) \sin^2{[\delta_{l+1}(E_k)-\delta_l(E_k)]},
\end{equation}
where $\delta_l(E_k)$ is the phase-shift for a given partial wave $l$ and collision energy $E_k$.

In this work, the phase-shifts for $^{174}$Yb$^+$-Li and $^{174}$Yb$^+$-Rb collisions from 1$\mu $K to 1K have been calculated using a single channel description of the scattering. In particular, the Numerov method is employed to propagate the wave function from a distance between $4.8\,a_{0}$ and $12\,000\,a_{0}$ with a step size of $0.006\,a_{0}$ for $^{174}$Yb$^+$-Li, whereas for $^{174}$Yb$^+$-Rb the propagation took place between $5.8\,a_{0}$ and $15\,000\,a_{0}$ with a step size of $0.003\,a_0$. The number of partial waves included varied with the collision energy, but we included as many as necessary to ensure a convergence better than $1\%$ of the elastic cross section.

Scattering properties at low collision energies are mainly dominated by the long-range tail of the atom-ion interaction potential. Therefore, we employ atom-ion potentials with the physical long-range but with an artificial short-range. In particular for $^{174}$Yb$^+$-Li the potential reads 

\begin{equation}
V(r)=\frac{C_6}{r^6}-\frac{C_4}{r^4},    
\end{equation}
with $C_4=82$ a.u. and $C_6=29284$ a.u. The same potential has been used previously in quasi-classical trajectory calculations showing that the short-range part of the atom-ion interactions has a negligible impact on scattering observables in the cold regime~\cite{Hirzler2020, Hirzler2022}. Whereas, for $^{174}$Yb$^+$-Rb the potential is taken as

\begin{equation}
V(r)=-\frac{C_4}{r^4}\left[1-\frac{1}{2}\left(\frac{r_m}{r}\right)^4 \right],    
\end{equation}
with $C_4=160$ and $r_m=10.142\,$a$_0$ corresponding to the a$^3\Sigma$ state. 

\section{ Numerical solution of the Generalized Langevin Equation}
\label{appendixB}

The description of the generalized Langevin dynamics of the ion in a bath requires to solve a set of stochastic integro-differential equations including (\ref{GLE}) and the equation associated to the generation of the noise. For the OU noise considered here the equations are 

\begin{equation}
\frac{\text{d}^{2}x}{\text{d}t^{2}} + \int_{0}^{t}\frac{\Gamma(t-t^{\prime})}{m} v(t^{\prime})\text{d}t^{\prime}+\frac{\Omega^{2}_{\text{RF}}}{4}[a + 2q\cos(\Omega_{\text{RF}}t)] x
= F_{x} 
\label{A1}
\end{equation}
\begin{eqnarray}
\frac{\text{d}x}{\text{d}t} = v \label{A2} \;\;\;\;\;\;\;\;\;\;\;\;\;\; \\  
\frac{\text{d} \zeta_{c}}{\text{d}t} = -\frac{1}{\tau_{\text{c}}}\zeta_{c} + \frac{\sqrt{D}}{\tau_{\text{c}}}\zeta \label{A3},
\end{eqnarray}
\noindent
where $F_{x} = \tilde{E}_{x}  + \zeta_{\text{c}}(t)$. The third equation is the OU process, $\zeta_{\text{c}}(t)$ refers to the generated colored noise and $\zeta(t)$ is the driver white noise. To solve numerically this set of equations we first solve the independent OU equation (\ref{A3}) using an stochastic finite difference scheme to generate a colored noise vector in the time grid. This noise vector is implemented in the solution of the two first coupled equations (\ref{A1}) and (\ref{A2}).

\begin{figure}[]
\begin{subfigure}[h]{1.0\linewidth}
\includegraphics[width=\linewidth]{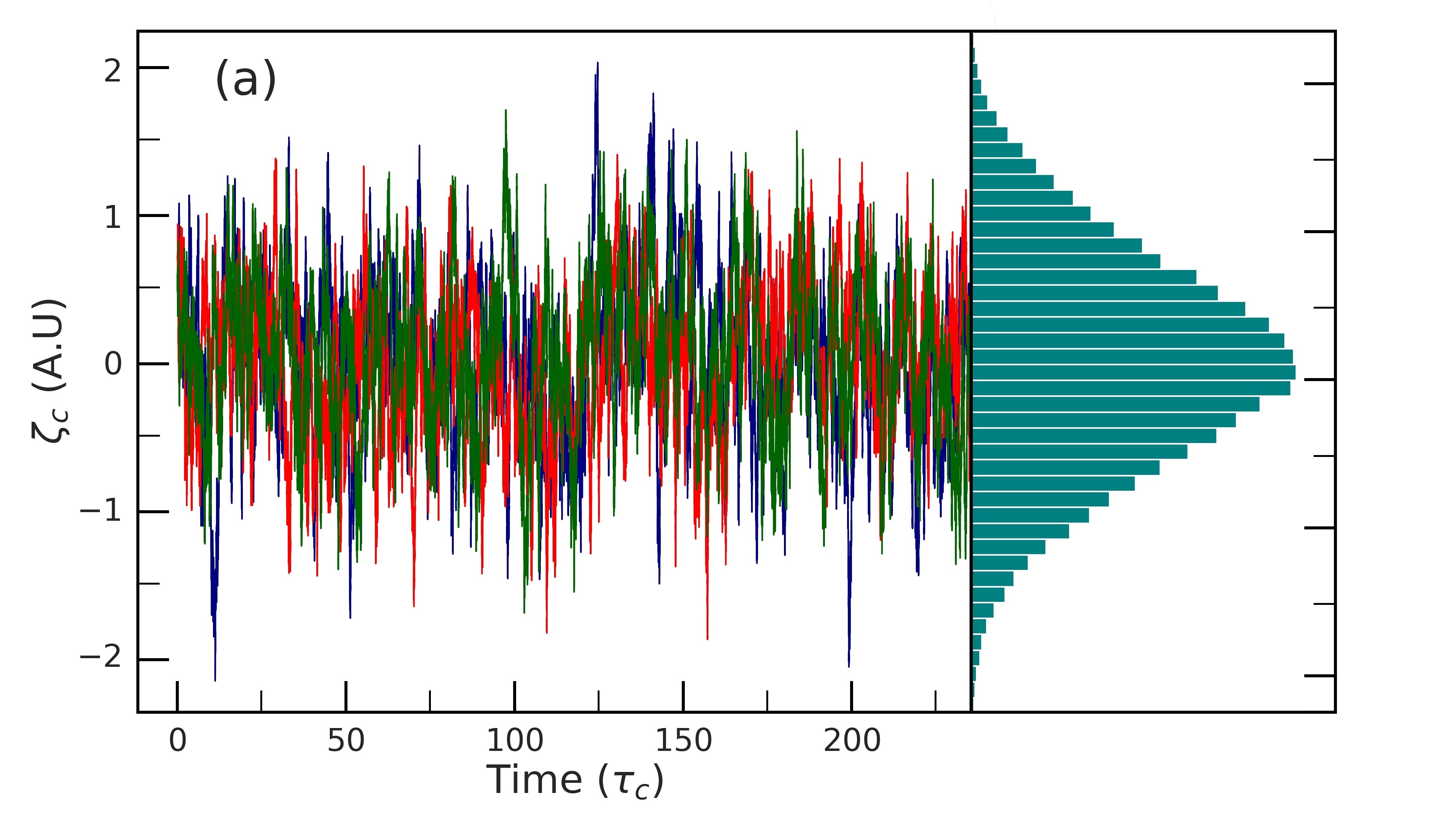}
\label{good_dist_agreement}
\end{subfigure}
\begin{subfigure}[h]{0.8\linewidth}
\includegraphics[width=\linewidth]{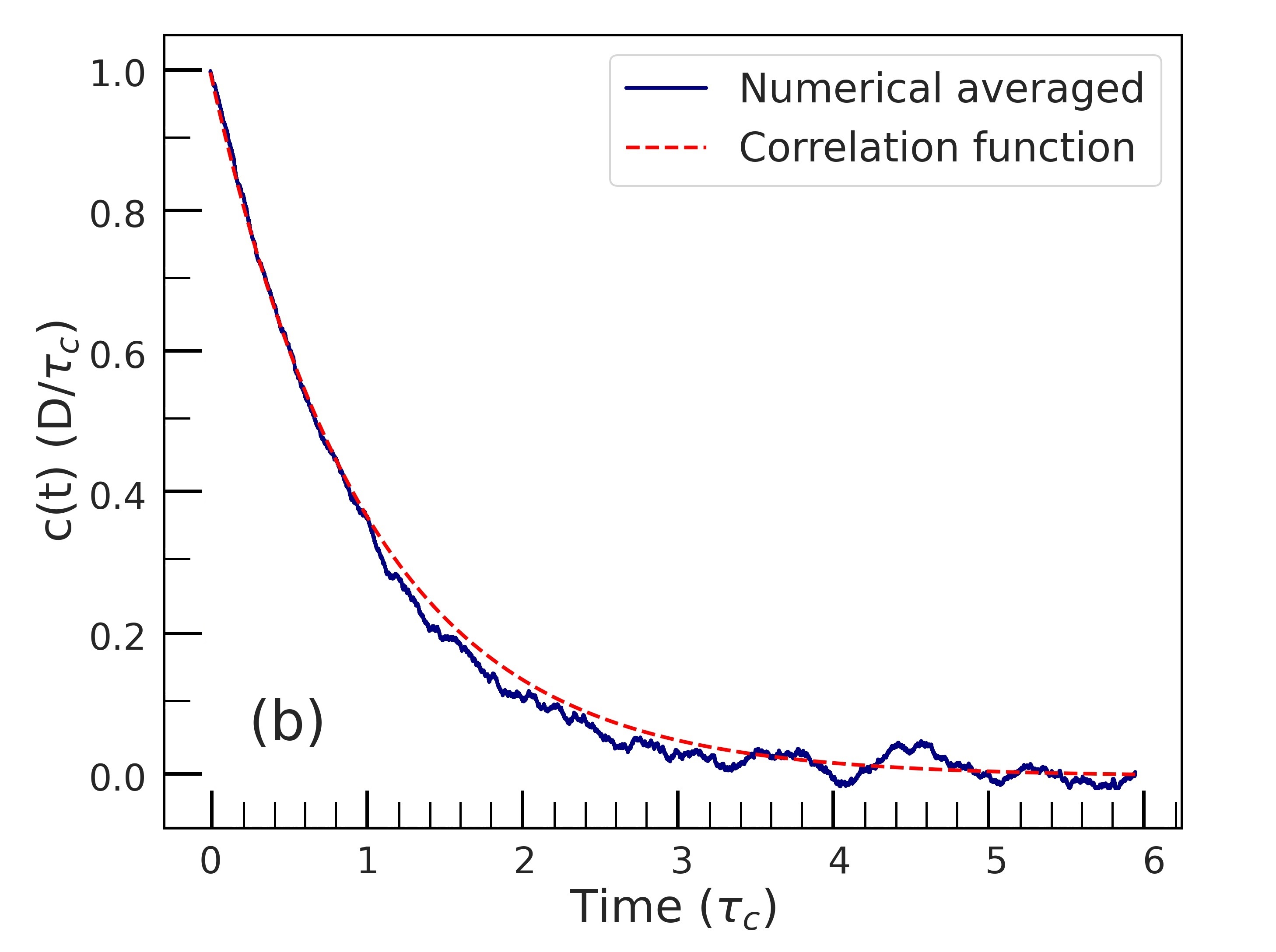}
\label{bad_dist_agreement}
\end{subfigure}%
 \caption{Colored Noise resulting from solving the OU equation (\ref{A3}). Panel (a) shows three different realizations of the OU noise and its distribution, panel (b) displays the correlation function of the noise fitting with Eq. (\ref{correlation_func}) }
\label{colored_noise}
\end{figure}

The finite difference scheme for the solution of Eq.~(\ref{A3}) is based on the method proposed in Ref.~\cite{finittedif}. The colored noise for each iteration is given by
\begin{equation*}
    \zeta_{\text{c}\,i+1} = \bigg( 1 - \frac{\Delta t}{\tau_{\text{c}}} \bigg)\zeta_{\text{c}\,i} + \frac{\sqrt{D}}{\tau_{\text{c}}}w(0,1)\sqrt{\Delta t},
\end{equation*}
where $\Delta t$ is the time step and $w(0,1)\sim N(0,1)$ is a Gaussian random variable with mean 0 and standard deviation equal to 1. The results are shown in Fig.~\ref{colored_noise}, where different realizations of the colored noise and its correlation function are displayed. 

Next, the solutions $\zeta_{\text{c} i}$ are used in the two coupled differential equations (\ref{A1}) and (\ref{A2}) for $r_{i}$ and $v_{i}$, which are solved using the following finite difference representation:
\begin{equation*}
\begin{split}
    x_{i+1} = &\bigg[ 2 - \frac{\Omega^{2}}{{4}}(a + 2q\cos(\Omega_{\rm{RF}}t))\Delta t^{2} \bigg]x_{i}\\
    -&x_{i-1}+ \bigg[\tilde{E}_{x} + \frac{\zeta_{c}(t)}{m} - S_{i}\bigg]\Delta t^{2},
    \end{split}
\end{equation*}
where $S_{i} = S(t_{i})$ is the integral term in Eq.~(\ref{GLE}). The correlation function and the fluctuation-dissipation relation allow us to express the integral in a recursive form as
\begin{equation*}
    S_{i+1} = S_{i} + \frac{1}{2k_{\rm{B}}T}\bigg( \exp\bigg(\frac{\Delta t}{\tau_{\rm{c}}} \bigg)v_{i+1} + v_{i}  \bigg).
\end{equation*}
Finally, the velocity vector is obtained using central differences. The procedure is repeated for several realizations $N_{\text{int}}$ ($\sim$ 10$^{6}$), and then the statistical average of the quantity (position or velocity) is computed to obtain its mean value.

In the case of a white noise, the Langevin dynamics is obtained after solving only Eq.~(\ref{LE}) using a finite difference scheme, yielding


\begin{widetext}
\begin{equation*}
    x_{i+1} = \frac{ \Big(\frac{\gamma^{\prime}\Delta t}{2}-1\Big)x_{i-1} - \Big(\frac{\Omega^{2}}{4}[a + 2q\cos(\Omega t)]\Delta t^{2} - 2\Big)x_{i} + \Big(\tilde{E}_{x} + \frac{\sqrt{D}w(0,1)}{m} \Big)\Delta t^{2}}{1+\frac{\gamma^{\prime} \Delta t}{2}},
\end{equation*}
\end{widetext}

and velocity is computed from central differences. 

The time scales associated with the Langevin dynamics of the ion are the Paul trap period (T$_{\text{RF}}$ = $2\pi$/$\Omega$), the relaxation time regarding the friction ($\tau_{\text{R}} = 1/\gamma^{\prime}$) and the correlation time ($\tau_{\rm{c}}$) for the case of colored Noise. For the atomic densities and bath temperatures considered in our calculations $\tau_{\rm{R}}$ and $\tau_{\rm{c}}$ are $\gtrsim$ 2$\times$10$^{-6}\,\text{s}$. Then the short time involved comes from the trap period (T$_{\text{RF}} \leq 1\times10^{-6}\text{s}$ ). Then, for the numerical solution we fix a time interval of $\Delta t = 5 \times 10^{-9}\,\text{s}$ which allow us to compute several steps within the same RF-oscillation. 

\section{Radial distribution function}

Using the approximate position distribution for the $x$ and $y$ component we find the radial distribution as
\begin{equation}
\begin{split}
    &\overline{P(r)} \propto  \, \, \, \, _{0}\tilde{F}_{1}(;1;\frac{\tilde{E}_{x}^{2}r^{2}}{\omega^{4}}) r \exp \Big(\frac{-m\omega^2r^{2}}{2k_{\rm{B}}T}\Big)\\
    & \xrightarrow[E_{x} = 0]{\text{}} \approx\frac{m\omega^{2}}{ k_{\rm{B}}T}r\exp \Big(\frac{-m\omega^2r^{2} }{2k_{\rm{B}}T}\Big), 
\end{split}
\label{radial_distribution}
\end{equation}

\noindent
where $_{0}\tilde{F}_{1}(;1;\frac{\tilde{E}_{x}^{2}r^{2}}{\omega^{4}})$ is the regularized hypergeometric function defined as
\begin{equation*}
    _{0}\tilde{F}_{1}(;1;\frac{\tilde{E}_{x}^{2}r^{2}}{\omega^{4}}) = \frac{1}{2\pi}\int_{0}^{2\pi}\exp\Big(\frac{2\tilde{E}_{x}r}{\omega^{2}}(\cos(\theta)+ \sin(\theta)) \Big)d\theta.
\end{equation*}

\noindent
In addition, the stationary distribution for the harmonic $z$-component is simply given by
\begin{equation}
        P(z) = \sqrt{\frac{m \omega_{z}^{2}}{2\pi k_{\text{B}}T}}\exp \Bigg[\frac{-m\omega_{z}^2 z^{2} }{2k_{\text{B}}T}\Bigg].
\end{equation}
where $\omega_{z}^{2} =\frac{\Omega_{\rm{RF}^{2}}}{4}a_{z}$.

\begin{figure}[]
\centering
\hspace{-0.1cm}
\includegraphics[scale=0.32
]{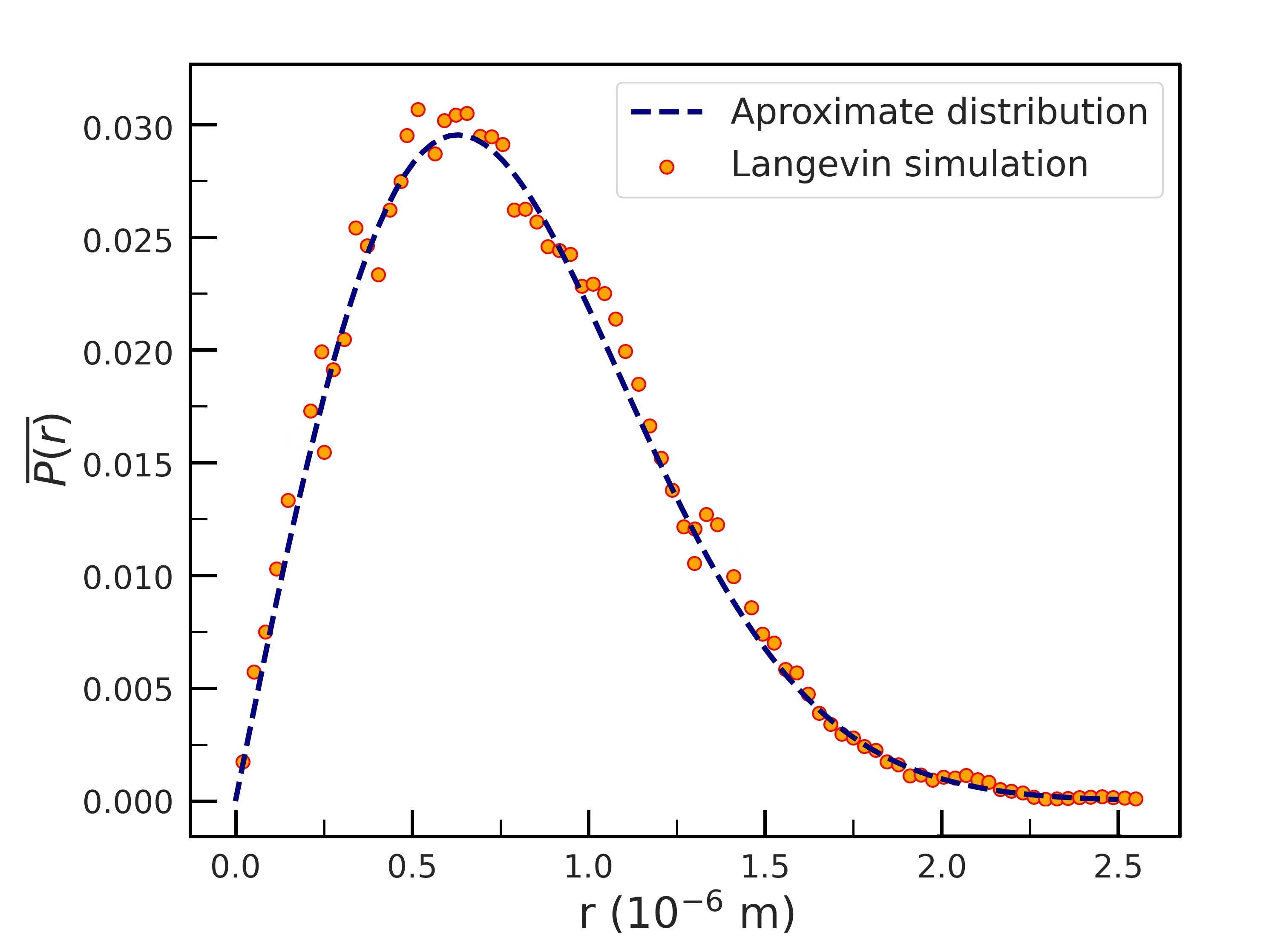}
\caption{Radial distribution of the ion fitting by the approximate  distribution (\ref{radial_distribution}). To achieved an accurate fitting the trap parameters are the same as those in fig. (\ref{fig4})}  
\label{radial_distribution_fig}
\end{figure}

Finally, $\overline{{P}(r)}$ and $P(z)$ give an idea of the averaged 3D ion position distribution in the trap. Fig.~(\ref{radial_distribution_fig}) shows the performance of the approximate radial distribution versus the numerical one. In the absence of excess micromotion, the regularized hypergeometric function equals 1. As a result, the distribution is identical to the radial distribution for a particle in the harmonic pseudopotential.

\newpage
\bibliography{apssamp}

\end{document}